\documentclass[conference]{IEEEtran}
\usepackage{textalpha}

\usepackage[T1]{fontenc}
\usepackage{lmodern}

\usepackage[utf8]{inputenc}
\usepackage{newtxtext}
\usepackage{tikz}
\usepackage{textgreek} 
\usetikzlibrary{positioning,arrows.meta,shapes.multipart,fit,calc,shapes.geometric,shapes,shapes.symbols,arrows}
\usetikzlibrary{shapes.geometric}
\usetikzlibrary{shapes.geometric}
\usepackage{tabularx}
\usepackage{booktabs}
\usepackage{siunitx}
\usepackage{tikz}
\usetikzlibrary{positioning,arrows.meta,shapes.multipart,fit,calc}
\usepackage{booktabs}
\usepackage{siunitx} 

\usepackage{longtable}
\usepackage{tabularx}
\usepackage{array}
\usepackage{graphicx}
\usepackage{amsmath}
\usepackage{url}
\usepackage{tikz}
\usetikzlibrary{arrows.meta, positioning, shapes.geometric, shapes, arrows, calc, fit, shapes.symbols}
\usepackage[backend=biber]{biblatex}

\usepackage[backend=biber]{biblatex}
\begin{document}

\title{Virginia Tech Transportation Safety Index (VTTSI)}

%
%
\author{\IEEEauthorblockN{Jason Cusati}
\IEEEauthorblockA{\textit{Computer Science} \\
Virginia Tech \\ Blacksburg, VA, USA \\
djjay@vt.edu}
\and
\IEEEauthorblockN{Cheng-Shun Chuang}
\IEEEauthorblockA{ \textit{Computer Science} \\
Virginia Tech \\ Alexandria, VA, USA \\
cchengshun@vt.edu}}

\maketitle

\sloppy
\raggedbottom

\begin{abstract}
The Virginia Tech Transportation Safety Index (VTTSI) is a real-time, cloud-native
framework for quantifying intersection safety using multimodal connected-vehicle
telemetry and multi-year VDOT crash history. Traditional crash-based methods rely on
lagged, aggregated data and cannot reflect rapidly changing operational conditions.
VTTSI addresses this gap through a hybrid modeling approach that fuses Empirical Bayes
(EB) crash stabilization, uplift factors derived from speed and conflict behavior, and a
CRITIC-weighted multi-criteria decision-making (MCDM) module combining SAW, EDAS, and
CODAS. The system produces interpretable, exposure-adjusted safety scores on a 0--100
scale every 15 minutes.

A cloud-deployed architecture built on FastAPI, PostgreSQL, PostGIS, and Streamlit
supports interactive visualization of traffic volumes, VRU exposure, speed variance, and
real-time incident activity. Validation across intersections demonstrates coherent
diurnal patterns, consistency among MCDM methods, and sensitivity to observable
operational turbulence. Sensitivity analysis further shows that the RT--SI is robust to
parameter perturbations, with deviations typically remaining below one point on the
0--100 scale.

By integrating long-term crash risk with short-term behavioral dynamics, VTTSI provides a
transparent, adaptive, and proactive safety-monitoring framework suitable for
transportation agencies, traffic management centers, fleet operators, and autonomous
vehicle systems.
\end{abstract}

\section{Introduction}

Assessing roadway safety has traditionally relied on retrospective analysis of
multi-year crash records obtained from transportation agencies such as VDOT. While
these crash-based approaches remain foundational for long-term planning,
they are inherently lagged and cannot capture rapidly changing operational
conditions at intersections. Recent studies, including work by Pribadi et al.\ and
Wicaksono et al.\ \cite{pribadi2022googlemapsfmea, wicaksono2022hiragooglemaps},
demonstrate that commercial routing platforms such as Google Maps and Apple Maps
optimize primarily for travel time rather than safety, and therefore do not provide
safety-aware routing or incorporate proactive risk measures.

Similarly, many vendor-provided intersection monitoring platforms compute
performance or safety scores using proprietary, black-box algorithms that rely on
limited operational indicators such as queue length, delay, or signal performance.
Although such systems provide mobility insight, they do not integrate long-term
crash history with real-time exposure or multimodal behavioral dynamics, nor do they
offer transparent or interpretable reasoning about safety conditions
\cite{schultz2025_surrogate}.

At the same time, modern connected-vehicle telemetry, IoT sensing, and roadside
monitoring systems are generating high-frequency multimodal data streams at a scale
that enables real-time safety assessment. Prior research in adaptive traffic
management has shown that integrating real-time sensor data with agent-based or
machine-learning-driven reasoning can improve congestion management and mobility
outcomes \cite{mutambik2025_iot_sensors}. However, analogous advances in real-time
safety assessment remain limited, especially in frameworks that explicitly combine
historical crash risk with emerging indicators such as speed variance, VRU
interactions, and conflict-like events.

These gaps motivate the need for a transparent, interpretable, and data-driven
real-time safety index that unifies long-term crash trends with short-term
operational signals. To address this need, the Virginia Tech Transportation Safety
Index (VTTSI) developed in this work fuses Empirical Bayes–stabilized crash risk with
heterogeneous real-time features, including speed distributions, multimodal exposure,
and automatically detected safety events. A parallel Multi-Criteria Decision-Making
(MCDM) module with CRITIC-derived weights provides an independent operational risk
ranking across intersections. Together, these components produce a 0--100 real-time
Safety Index updated every 15 minutes and presented through a cloud-native
architecture.

The design, validation, and analysis of VTTSI are organized around four research
questions that frame the contributions of the system and the safety-related insights
the framework enables.

\subsection*{Research Questions}

\noindent\textbf{RQ1:} \emph{Can multi-year crash history and real-time connected-vehicle telemetry be combined into a unified, exposure-adjusted, and interpretable real-time safety index?} 

This question examines whether long-term statistical risk (via Empirical Bayes) and
short-term operational turbulence (via uplift factors and MCDM criteria) can coexist
meaningfully in a single safety metric.

\noindent\textbf{RQ2:} \emph{Which real-time features most strongly influence short-term
intersection risk, and how sensitive is the Safety Index to changes in these
features?}  
This includes analysis of speed variance, VRU exposure, conflict events, and volume
patterns, and whether these signals provide stable real-time indicators of elevated
risk.

\noindent\textbf{RQ3:} \emph{Does a hybrid crash-based and MCDM-based safety index produce 
consistent and interpretable temporal patterns across different intersections and
operating conditions?}  
This explores whether the two modeling pipelines align or diverge, and what their
agreement reveals about intersection behavior.

\noindent\textbf{RQ4:} \emph{What limitations arise from representing intersection safety with 
flat, tabular features, and how might relational or agentic approaches address these
limitations?}  
This question motivates the Agentic Traffic Safety Knowledge Graph (TS--KG)
introduced later in the paper.

These research questions provide a conceptual foundation for the analyses that
follow and help situate VTTSI within the broader landscape of transportation safety
research.

\section{Related Work}
\label{sec:relatedwork}

Safety assessment approaches can be grouped into four major areas:
(1) crash-based statistical methods,
(2) surrogate safety indicators,
(3) multi-criteria decision-making (MCDM) methods, and
(4) data-driven predictive models. Each contributes valuable insight while also
exhibiting limitations that VTTSI seeks to address.

\subsection{Crash-Based and Empirical Bayes Methods}

Crash-based network screening traditionally relies on crash frequencies, crash rates,
and expected-versus-observed comparisons. Persaud and Lyon \cite{persaud2007}
demonstrated that Empirical Bayes (EB) methods mitigate regression-to-the-mean
effects by combining observed crash counts with Safety Performance Functions (SPFs).
Systemic approaches further incorporate exposure and roadway features
\cite{montella2020systemic}. However, all such methods depend on multi-year crash
aggregates, which limits their responsiveness to evolving real-time operating
conditions.

\subsection{Surrogate Safety and Operational Indicators}

Surrogate safety indicators—including post-encroachment time (PET), speed variance,
and conflict counts—provide insight into short-term turbulence and emerging hazards.
Schultz et al.\ \cite{schultz2025_surrogate} showed that surrogate indicators capture
risk in locations with sparse crash data but often lack consistent aggregation frameworks
and may not reflect long-term crash likelihood. Without integration into a larger
modeling framework, these indicators can be noisy or difficult to interpret.

\subsection{Multi-Criteria Decision-Making in Transportation Safety}

MCDM methods such as SAW, EDAS \cite{EDAS}, and CODAS \cite{CODAS} combine heterogeneous
indicators into unified safety or performance scores. Recent work by Amraji et al.
\cite{Amraji2025CombinedSafetyIndex} demonstrated that CRITIC-derived objective weights
provide robust handling of correlated criteria in transportation datasets. However,
most existing applications rely on static, offline datasets, leaving their real-time
potential underexplored.

\subsection{Machine Learning and Data-Driven Approaches}

Data-driven models leverage continuous sensing, connected-vehicle telemetry, and
high-resolution trajectories to predict crash likelihood, detect conflict events, or
estimate real-time risk \cite{HUANG2020105392, Zhang31122025, Almutairi2025}. While
these methods can be highly predictive, many lack interpretable outputs suitable for
operational deployment or integration into agency workflows.

\subsection{Positioning of VTTSI}

VTTSI bridges these approaches by combining EB stabilized crash risk, real-time
uplift factors, and CRITIC-weighted MCDM scoring in a cloud-native environment. This
integration enables a unified safety index that reflects both long-term crash history
and short-term operational disturbance, addressing key limitations of crash-only,
surrogate-only, and ML-only methods.

\section{Dataset Description}
\label{sec:dataset}

The Real-Time Safety Index (RT--SI) and MCDM Safety Index are computed using 
data stored in a Cloud SQL PostgreSQL database populated by the VTTI Trino 
system. The backend services (FastAPI) query six primary real-time tables—
\texttt{bsm}, \texttt{psm}, \texttt{"vehicle-count"}, \texttt{"vru-count"}, 
\texttt{"speed-distribution"}, \texttt{"safety-event"}—together with a 
historical crash dataset derived from VDOT records.  
These datasets provide high-frequency telemetry, multimodal counts, speed 
distributions, safety events, and long-term crash trends, enabling 
15-minute real-time computation of intersection-level safety scores.

\subsection{Basic Safety Messages (BSM)}
The \texttt{bsm} table stores raw J2735 Basic Safety Messages transmitted by 
connected vehicles. Records include vehicle position, speed, heading, 
acceleration, yaw rate, brake status, and vehicle size.  
The backend aggregates these messages into 15-minute bins to estimate:
\begin{itemize}
    \item vehicle exposure,
    \item average speed,
    \item speed variance,
    \item turbulence and hard-braking indicators.
\end{itemize}

\subsection{Personal Safety Messages (PSM)}
The \texttt{psm} table provides telemetry from pedestrians and cyclists 
(VRUs), including position, movement type, device activity, and dynamic 
attributes such as acceleration and yaw rate. These records support VRU 
exposure estimation and vehicle--VRU conflict assessment.

\subsection{Vehicle and VRU Count Tables}
Two count tables supply approach-level exposure data:
\begin{itemize}
    \item \textbf{\texttt{"vehicle-count"}}: movement-level classified vehicle counts 
          (left, through, right) aggregated every 15 minutes.
    \item \textbf{\texttt{"vru-count"}}: pedestrian and bicycle volumes, also binned 
          at 15-minute intervals.
\end{itemize}

\subsection{Speed Distribution Table}
The \texttt{"speed-distribution"} table records the number of vehicles in 
predefined speed bins (e.g., \texttt{"0--5 mph"}, \texttt{"20--25 mph"}).  
The midpoint of each bin, weighted by event count, is used to compute:
\begin{itemize}
    \item average speed, and
    \item speed variance,
\end{itemize}
which form key components of real-time uplift factors and MCDM criteria.

\subsection{Safety Event Table}
The \texttt{"safety-event"} table contains automatically detected safety-critical 
events, such as red-light running, intersection conflicts (IC), lane-change 
violations (LCV), and near-miss detections.  
Incident counts aggregated per 15-minute bin are primary risk signals in both 
RT--SI and MCDM scoring.

\subsection{VDOT Historical Crash Dataset}
A separate offline historical dataset, referred to as \textbf{Dataset~1}, contains
all police-reported crashes from VDOT (2017--2024). Each crash record is spatially
joined to the nearest instrumented intersection using:
\begin{itemize}
    \item VDOT Linear Referencing System (LRS) centerline network, and
    \item road–intersection geospatial matching,
\end{itemize}
allowing each crash to be assigned to a specific intersection in the Smart 
Intersection system.

The table includes:
\begin{itemize}
    \item intersection identifier,
    \item crash date and time,
    \item severity (fatal, injury, PDO),
    \item injury counts (K, A, B),
    \item roadway, weather, and light conditions,
    \item collision type and work zone attributes.
\end{itemize}

These data are used to compute:
\begin{itemize}
    \item \textbf{historical baseline crash rates},
    \item \textbf{severity-weighted risk}, and
    \item the \textbf{Empirical Bayes prior} for stabilizing short-term risk.
\end{itemize}

\begin{table}[h]
\centering
\caption{Dataset~1: VDOT crash data used for historical baseline modeling}
\label{dataset:vdot}
\begin{tabular}{l}
\toprule
File: \texttt{vdot\_crash\_with\_intersections} \\
Content: Police-reported crashes (2017--2024) joined to intersections \\
\bottomrule
\end{tabular}
\end{table}

\subsection{Data Integration in the Backend}

Integration and aggregation are performed by two core backend components:
the \textbf{MCDM Safety Index Service} (implemented as 
\texttt{MCDMSafetyIndexService}) and the \textbf{RT–SI Safety Index Service} 
(implemented as \texttt{RTSISafetyIndexService}). These services orchestrate 
the retrieval, alignment, and transformation of real-time and historical data 
into the feature matrices used by the scoring algorithms.

Key steps include:
\begin{itemize}
    \item \textbf{Time binning:}
    all data sources are synchronized into 15-minute intervals using a unified 
    timestamp floor operation.
    \item \textbf{Cross-source alignment:}
    the most recent timestamp common to BSM, PSM, count tables, speed 
    distributions, and events is selected to avoid misalignment.
    \item \textbf{Feature extraction:}
    aggregated signals include vehicle exposure, VRU exposure, speed metrics, 
    safety-event frequency, and conflict-related indicators.
    \item \textbf{Crash-history fusion:}
    the RT–SI service combines Empirical Bayes–stabilized historical rates with 
    uplift factors.
    \item \textbf{Missing-data handling:}
    empty bins are imputed with zeros to ensure consistent matrix dimensions.
\end{itemize}

\section{Methodology}
\label{sec:methodology}

\begin{figure*}
    \centering
    \includegraphics[width=\linewidth]{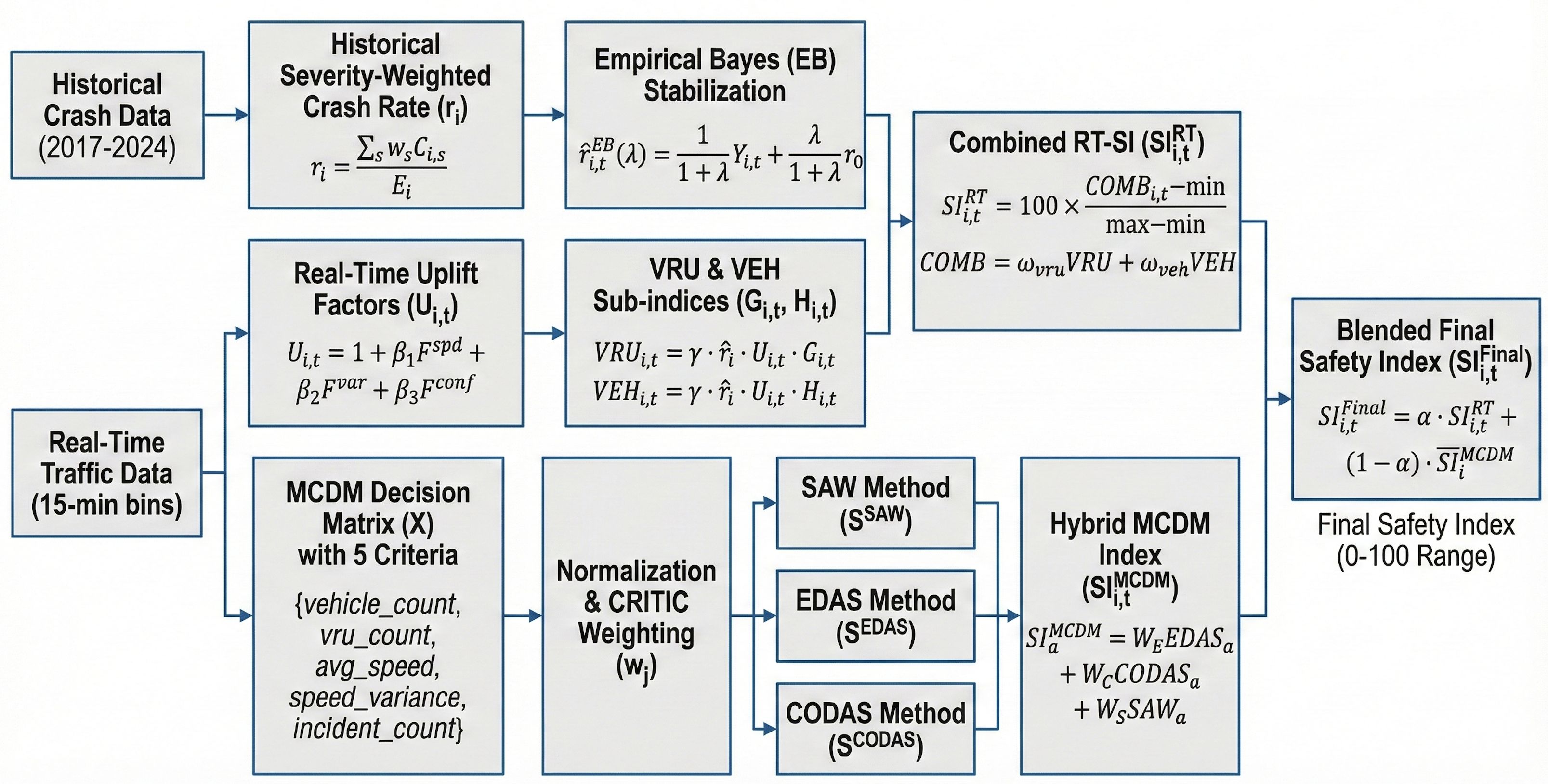}
    \caption{Methodology overview}
    \label{fig:methodology-pic}
\end{figure*}

The methodology has two parallel safety engines: (1) RT–SI captures historical crash risk adjusted by real-time uplift factors, and (2) MCDM captures instantaneous, anomaly-based operational risk. Both produce 0–100 indices which are later blended ($\alpha$-controlled) to form a single Final Safety Index. This section describes each component in the order they are computed in production.

\subsection{Real-Time Safety Index (RT-SI)}
We compute a per-intersection, per-15-minute Safety Index that is severity-weighted, exposure-adjusted, stabilized by Empirical Bayes (EB), and adjusted by real-time operating factors.

RT–SI represents a short-term operational safety estimate anchored by a long-term historical baseline. EB stabilization ensures crash-free intersections do not inflate risk spuriously. The uplift factors magnify baseline risk only when real-time indicators (speed deficit, variance, conflicts) deviate from expected safe operating conditions.

\paragraph{Historical severity-weighted crash rate.}
\begin{equation}
r_i \;=\; \frac{\sum_{s\in\{\text{Fatal,Injury,PDO}\}} w_s\,C_{i,s}}{E_i},
\end{equation}
where $C_{i,s}$ are crash counts of severity $s$, $w_s$ are severity weights (e.g., $w_F{=}10,\ w_I{=}3,\ w_P{=}1$), and $E_i$ is the exposure (e.g., VMT or entering vehicle volume).

\paragraph{Empirical Bayes stabilization (no-exposure case).}
Let $Y_{i,t}$ denote the severity-weighted crash count for site $i$ and
15-minute bin $t$. For the training period 2017--2024 we first compute the
global mean rate
\begin{equation}
r_0 \;=\; \frac{1}{N} \sum_{i,t \in \text{2017--2024}} Y_{i,t},
\end{equation}
where $N$ is the total number of $(i,t)$ combinations in 2017--2024.

Given a candidate shrinkage parameter $\lambda$, the Empirical Bayes (EB)
estimate for each site/bin is
\begin{equation}
\hat r^{\text{EB}}_{i,t}(\lambda)
\;=\;
\frac{1}{1+\lambda}\,Y_{i,t}
\;+\;
\frac{\lambda}{1+\lambda}\,r_0.
\end{equation}
Because we do not model exposure explicitly in this step, the EB rate is equal
to the expected count, i.e.
\begin{equation}
\hat Y_{i,t}^{(2025)}(\lambda)
\;=\;
\hat r^{\text{EB}}_{i,t}(\lambda).
\end{equation}

To choose $\lambda$, we treat 2017--2024 as training data and 2025 as a
hold-out set and minimize the Poisson log-loss
\begin{equation}
L(\lambda)
=
\sum_{i,t \in \text{2025}}
\Big(
\hat Y_{i,t}^{(2025)}(\lambda)
-
Y_{i,t}^{(2025)}
\log \hat Y_{i,t}^{(2025)}(\lambda)
\Big),
\end{equation}
over a log-spaced grid
\begin{equation}
\begin{aligned}
\lambda \in \{&
0.1,\ 0.3,\ 1,\ 3,\ 10,\ 30,\ 100, \\
&300,\ 1000,\ 10000,\ 30000,\ 100000\}.
\end{aligned}
\end{equation}

After extending this grid, the optimal value was found to be
\[
\lambda^\star = 100000.
\]
In production we fix $\lambda = \lambda^\star$ and use the corresponding
$\hat r^{\text{EB}}_{i,t}(\lambda^\star)$ as the stabilized historical component
in the Safety Index. This follows standard Empirical Bayes practice in roadway safety
modeling~\cite{persaud2007, montella2020systemic}.

\paragraph{Real-time uplift factors.}
\begin{align}
F^{\text{spd}}_{i,t} &= \min\!\left(1,\;k_1\frac{v^{\text{FF}}_i-\bar v_{i,t}}{v^{\text{FF}}_i}\right), \\
F^{\text{var}}_{i,t} &= \min\!\left(1,\;k_2\frac{\sigma_{v,i,t}}{\bar v_{i,t}+\varepsilon}\right), \\
F^{\text{conf}}_{i,t} &= \min\!\left(1,\;k_3\frac{\text{turningVol}_{i,t}\cdot V^{\text{vru}}_{i,t}}{\text{scale}}\right).
\end{align}
We combine these into an uplift factor:
\begin{equation}
U_{i,t} = 1 + \beta_1F^{\text{spd}}_{i,t} + \beta_2F^{\text{var}}_{i,t} + \beta_3F^{\text{conf}}_{i,t}.
\end{equation}

\paragraph{Sub-indices}
\begin{align}
G_{i,t} &= \min\!\left(1,\;k_4\frac{V^{\text{vru}}_{i,t}}{V^{\text{veh}}_{i,t}+\varepsilon}\right), \\
\mathrm{VRU}_{i,t} &= \gamma \cdot \hat r_i \cdot U_{i,t} \cdot G_{i,t}, \\
H_{i,t} &= \min\!\left(1,\;k_5\frac{V^{\text{veh}}_{i,t}}{\mathrm{capacity}_i}\right), \\
\mathrm{VEH}_{i,t} &= \gamma \cdot \hat r_i \cdot U_{i,t} \cdot H_{i,t}.
\end{align}

\paragraph{Combined and scaled index}
\begin{equation}
\mathrm{COMB}_{i,t} \;=\; \omega_{\text{vru}}\mathrm{VRU}_{i,t} + \omega_{\text{veh}}\mathrm{VEH}_{i,t},\quad
\omega_{\text{vru}}+\omega_{\text{veh}}=1,
\end{equation}
\begin{equation}
\mathrm{SI}^{\text{RT}}_{i,t} \;=\; 100 \times
\frac{\mathrm{COMB}_{i,t}-\min}{\max-\min}.
\end{equation}

\subsection{Decision Matrix and Temporal Aggregation}

For the Multi-Criteria Decision Making (MCDM) Safety Index, we construct a
decision matrix where each alternative corresponds to an
\emph{intersection--time-bin} pair and each criterion is a real-time traffic or
safety measure.

We use five criteria:
\begin{equation}
\begin{aligned}
\mathcal{C} = \{\, 
&\text{vehicle\_count},\ \text{vru\_count},\ \text{avg\_speed},\\
&\text{speed\_variance},\ \text{incident\_count}
\,\}.
\end{aligned}
\end{equation}
For a given evaluation window (e.g., the last 24 hours), we aggregate data into
15-minute time bins. Each row of the decision matrix $X$ is then
\[
a = (i,t),
\]
where $i$ denotes an intersection and $t$ a 15-minute time bin within the
lookback window. The entry $x_{a j}$ is the value of criterion $j \in \mathcal{C}$
for intersection $i$ in time bin $t$. Thus the MCDM weights explicitly depend on
both \emph{spatial} variation (across intersections) and \emph{temporal}
variation (across 15-minute bins in the lookback horizon).

We normalize each criterion via min--max scaling:
\begin{equation}
\tilde x_{a j}
=
\begin{cases}
\dfrac{x_{a j} - \min_a x_{a j}}{\max_a x_{a j} - \min_a x_{a j}}, & \text{if }\max_a x_{a j} > \min_a x_{a j},\\[0.8em]
0, & \text{otherwise}.
\end{cases}
\end{equation}
The resulting matrix $\tilde X$ is used for all subsequent MCDM steps.

\subsection{CRITIC-Based Criterion Weights}

We use the CRITIC (CRiteria Importance Through Intercriteria Correlation) method
to derive objective weights for the five criteria based on their variability and
mutual correlation over all intersection--time alternatives in the lookback
window~\cite{Amraji2025CombinedSafetyIndex}.

Let $\tilde x_{\cdot j}$ be the $j$th normalized column and $\sigma_j$ its
standard deviation:
\[
\sigma_j = \mathrm{sd}(\tilde x_{\cdot j}).
\]
Let $\rho_{jk}$ denote the Pearson correlation between criteria $j$ and $k$
over all rows. The conflict (or contrast) of criterion $j$ is
\begin{equation}
\Gamma_j = \sum_{k=1}^m (1 - \rho_{jk}),
\end{equation}
where $m = |\mathcal{C}| = 5$. The information content of criterion $j$ is
\begin{equation}
I_j = \sigma_j \Gamma_j,
\end{equation}
and the CRITIC weight is
\begin{equation}
w_j =
\begin{cases}
\dfrac{I_j}{\sum_{k=1}^m I_k}, & \text{if }\sum_k I_k > 0,\\[0.8em]
\dfrac{1}{m}, & \text{otherwise}.
\end{cases}
\end{equation}
These weights are recomputed dynamically using the last 24 hours of
intersection--time-bin data (or 1-day lookback around the target time for
intersection-specific queries), ensuring that the MCDM Safety Index reflects
recent spatial--temporal patterns in volumes, speeds, and incidents, following
CRITIC-based weighting practice in transportation MCDM studies~\cite{Amraji2025CombinedSafetyIndex}.

\subsubsection{Rationale for CRITIC Weighting}

We use CRITIC because it derives data-driven criterion weights from variability and correlation, down-weighting redundant indicators and up-weighting informative ones. Since weights are recalculated over the last 24 hours, the MCDM index automatically adapts to changing conditions without manual tuning.

\subsection{SAW (Simple Additive Weighting)}

Given the normalized matrix $\tilde X$ and CRITIC weights $\{w_j\}$, the SAW
score for alternative $a$ is
\begin{equation}
S^{\text{SAW}}_a = \sum_{j \in \mathcal{C}} w_j \tilde x_{a j}.
\end{equation}
We then scale SAW scores to a 0--100 range:
\begin{equation}
\mathrm{SAW}_a =
\begin{cases}
100 \displaystyle
\frac{
S^{\text{SAW}}_a - \min_a S^{\text{SAW}}_a
}{
\max_a S^{\text{SAW}}_a - \min_a S^{\text{SAW}}_a
},
& \\
\quad\text{if }\max_a S^{\text{SAW}}_a > \min_a S^{\text{SAW}}_a, \\[0.6em]
50, & \text{otherwise}.
\end{cases}
\end{equation}
This implementation follows the standard SAW formulation as used in transportation
MCDM applications~\cite{Amraji2025CombinedSafetyIndex}.

\subsection{EDAS (Evaluation Based on Distance from Average Solution)}

For EDAS, we first compute the average solution for each criterion:
\begin{equation}
\bar x_j = \frac{1}{N} \sum_a \tilde x_{a j},
\end{equation}
where $N$ is the number of intersection--time alternatives.

The positive and negative distances from the average solution are
\begin{align}
\mathrm{PDA}_{a j} &= \max\!\left(0,\ \frac{\tilde x_{a j} - \bar x_j}{\bar x_j}\right),\\
\mathrm{NDA}_{a j} &= \max\!\left(0,\ \frac{\bar x_j - \tilde x_{a j}}{\bar x_j}\right).
\end{align}
Weighted sums of the distances are
\begin{align}
\mathrm{SP}_a &= \sum_{j \in \mathcal{C}} w_j\,\mathrm{PDA}_{a j},\\
\mathrm{SN}_a &= \sum_{j \in \mathcal{C}} w_j\,\mathrm{NDA}_{a j}.
\end{align}
We normalize these as
\begin{align}
\mathrm{SP}^{\text{norm}}_a &= \frac{\mathrm{SP}_a}{\max_a \mathrm{SP}_a} \quad\text{(if }\max_a \mathrm{SP}_a > 0\text{)},\\
\mathrm{SN}^{\text{norm}}_a &= \frac{\mathrm{SN}_a}{\max_a \mathrm{SN}_a} \quad\text{(if }\max_a \mathrm{SN}_a > 0\text{)},
\end{align}
and define the EDAS appraisal score
\begin{equation}
S^{\text{EDAS}}_a = \frac{1}{2}\left(\mathrm{SP}^{\text{norm}}_a + \left(1 - \mathrm{SN}^{\text{norm}}_a\right)\right).
\end{equation}
Finally, we scale to 0--100:
Let
$S^{\min} = \min_a S^{\mathrm{EDAS}}_a$
and
$S^{\max} = \max_a S^{\mathrm{EDAS}}_a$.
\begin{equation}
\mathrm{EDAS}_a =
\begin{cases}
\displaystyle
100\,\dfrac{S^{\mathrm{EDAS}}_a - S^{\min}}{S^{\max} - S^{\min}},
& \text{if } S^{\max} > S^{\min},\\[0.6em]
50, & \text{otherwise.}
\end{cases}
\end{equation}
The above steps follow the canonical EDAS procedure~\cite{EDAS}.

\subsection{CODAS (Combinative Distance-Based Assessment)}

For CODAS, we first create a weighted normalized matrix
\begin{equation}
z_{a j} = w_j \tilde x_{a j}.
\end{equation}
The negative-ideal solution (NIS) is the column-wise minimum:
\begin{equation}
z^-_j = \min_a z_{a j}.
\end{equation}
For each alternative $a$, we compute the Euclidean and Taxicab distances from
the NIS:
\begin{align}
E_a &= \sqrt{\sum_{j \in \mathcal{C}} (z_{a j} - z^-_j)^2},\\
T_a &= \sum_{j \in \mathcal{C}} |z_{a j} - z^-_j|.
\end{align}
CODAS constructs a pairwise relative assessment matrix
\begin{equation}
\Psi_{ab} = 
\begin{cases}
E_a - E_b, & \text{if }E_a - E_b \neq 0,\\
T_a - T_b, & \text{if }E_a - E_b = 0.
\end{cases}
\end{equation}
The CODAS appraisal score is
\begin{equation}
S^{\text{CODAS}}_a = \sum_b \Psi_{ab},
\end{equation}
which we again scale to 0--100:
\begin{equation}
\mathrm{CODAS}_a =
\begin{cases}
\displaystyle
100 \frac{
S^{\mathrm{CODAS}}_a - \min S
}{
\max S - \min S
}, \\[0.8em]
50,
\end{cases}
\qquad
\begin{aligned}
&(\Delta S > 0),\\
&\text{else}.
\end{aligned}
\end{equation}
This CODAS implementation follows the original formulation~\cite{CODAS}.

\subsection{Hybrid MCDM Index with CRITIC Method Weights}

For each intersection--time alternative $a$ we thus obtain three method scores:
\[
\mathrm{EDAS}_a,\quad \mathrm{CODAS}_a,\quad \mathrm{SAW}_a,
\]
each in the 0--100 range (higher = higher risk).

To aggregate them, we again use the CRITIC method---this time with the
\emph{methods} as criteria and the intersection--time alternatives as “samples”.
We form a $N \times 3$ matrix
\[
M = 
\big[\,
\mathrm{EDAS}_a,\ \mathrm{CODAS}_a,\ \mathrm{SAW}_a
\,\big]_{a=1}^N
\]
and compute method-level CRITIC weights
\[
\{W_E,\ W_C,\ W_S\}
\]
using the same standard-deviation and correlation-based procedure as for the
criteria weights.

The hybrid MCDM Safety Index for alternative $a = (i,t)$ is then
\begin{equation}
\begin{aligned}
\mathrm{SI}^{\mathrm{MCDM}}_a
&=
W_E \,\mathrm{EDAS}_a
+
W_C \,\mathrm{CODAS}_a
+
W_S \,\mathrm{SAW}_a,
\\[4pt]
&\qquad a = 1,\ldots,N.
\end{aligned}
\end{equation}
\begin{equation}
\overline{\mathrm{SI}}^{\mathrm{MCDM}}
=
\frac{1}{N}
\sum_{a=1}^{N}
\mathrm{SI}^{\mathrm{MCDM}}_a
=
\frac{1}{N}\,\mathbf{1}^{\mathsf{T}} M \mathbf{W}.
\end{equation}

with $W_E + W_C + W_S = 1$. In the implementation, we set the
\emph{Safety Score} equal to this hybrid index (no inversion), so higher values
correspond directly to higher estimated risk for that intersection and
15-minute time bin, consistent with hybrid MCDM formulations in transportation
safety~\cite{Amraji2025CombinedSafetyIndex}.

\subsection{Blended Final Index}
To harmonize real-time safety with long-term prioritization, we define
\begin{equation}
\label{eq:blended-final}
\mathrm{SI}^{\text{Final}}_{i,t} = \alpha \cdot \mathrm{SI}^{\text{RT}}_{i,t} +
(1-\alpha)\cdot \overline{\mathrm{SI}}^{\text{MCDM}}_i,
\end{equation}
where $\alpha$ tunes the emphasis (e.g., $\alpha=0.7$ for driver-facing dashboards).

\subsubsection{Rationale for a Blended Safety Index}

We blend the real-time MCDM with the historically informed RT–SI index because each captures complementary aspects of risk: RT–SI reflects immediate operational turbulence, while the MCDM index integrates long-term crash patterns and structural exposure. The weighted combination reduces volatility, prevents overreaction to noisy inputs, and produces a more stable and interpretable score for deployment~\cite{montella2020systemic, Amraji2025CombinedSafetyIndex}. Full justification is provided in the Appendix.

\section{System Design}
\label{sec:systemdesign}

\begin{figure*}[t]
\centering
\resizebox{\textwidth}{!}{%
\begin{tikzpicture}[
    >=Latex,
    font=\small,
    user/.style    ={rectangle, draw, rounded corners=4pt, fill=gray!12,
                     minimum width=15cm, minimum height=0.85cm, align=center},
    page/.style    ={rectangle, draw, rounded corners=3pt, fill=blue!9,
                     minimum width=2.2cm, minimum height=1.35cm,
                     align=center, font=\scriptsize},
    chatpage/.style={rectangle, draw, rounded corners=3pt, fill=orange!22,
                     minimum width=2.2cm, minimum height=1.35cm,
                     align=center, font=\scriptsize},
    api/.style     ={rectangle, draw, rounded corners=3pt, fill=green!9,
                     minimum width=7.8cm, minimum height=1.1cm,
                     align=center, font=\scriptsize},
    chatapi/.style ={rectangle, draw, rounded corners=3pt, fill=orange!14,
                     minimum width=5.0cm, minimum height=1.1cm,
                     align=center, font=\scriptsize},
    svc/.style     ={rectangle, draw, rounded corners=3pt, fill=yellow!14,
                     minimum width=4.0cm, minimum height=1.7cm,
                     align=center, font=\scriptsize},
    chatsvc/.style ={rectangle, draw, rounded corners=3pt, fill=orange!18,
                     minimum width=4.0cm, minimum height=1.7cm,
                     align=center, font=\scriptsize},
    llm/.style     ={rectangle, draw, rounded corners=3pt, fill=purple!12,
                     minimum width=2.8cm, minimum height=1.7cm,
                     align=center, font=\scriptsize},
    dbtab/.style   ={rectangle, draw, rounded corners=3pt, fill=red!7,
                     minimum width=6.0cm, minimum height=1.3cm,
                     align=center, font=\scriptsize},
    ext/.style     ={rectangle, draw, dashed, rounded corners=3pt, fill=gray!7,
                     minimum width=5.8cm, minimum height=1.0cm,
                     align=center, font=\scriptsize},
    grpbox/.style  ={rectangle, draw, dashed, rounded corners=5pt, inner sep=5pt},
    arr/.style     ={->, thick, >=Latex},
    darr/.style    ={->, thick, dashed, >=Latex},
]

\node[user] (user)
    {\textbf{End Users}
     \enspace(Web Browsers \ | \ Transportation Agencies \ | \
              Traffic Management Centers)};

\node[page, below=1.3cm of user, xshift=-5.8cm] (p0)
    {\textbf{Operations}\\KPIs + Map\\$\alpha$-Blend};
\node[page, right=0.2cm of p0] (p1)
    {\textbf{Trend}\\Analysis\\Time-Series};
\node[page, right=0.2cm of p1] (p2)
    {\textbf{Analytics}\\Validation\\Correlations};
\node[page, right=0.2cm of p2] (p3)
    {\textbf{Sensitivity}\\Analysis\\Perturbation};
\node[page, right=0.2cm of p3] (p4)
    {\textbf{Database}\\Explorer\\Read-Only SQL};
\node[chatpage, right=0.6cm of p4] (p5)
    {\textbf{SafetyChat}\\Persistent\\Chat Dock};

\node[grpbox, fit=(p0)(p1)(p2)(p3)(p4)(p5),
      label={[font=\small\bfseries, fill=white, inner sep=2pt,
              rounded corners=2pt]above:{Vite React Dashboard (Google Cloud Run)}}]
      (fe_box) {};

\node[api, below=1.6cm of p1, xshift=0.4cm] (api_main)
    {\textbf{FastAPI --- Safety REST API}\quad\textit{(Google Cloud Run)}\\[2pt]
     \texttt{/api/v1/safety/index} \quad
     \texttt{/api/v1/analytics} \quad
     \texttt{/api/v1/history} \quad
     \texttt{/api/v1/database\_explorer}};

\node[chatapi, right=0.5cm of api_main] (api_chat)
    {\textbf{SafetyChat API}\\
     \textit{(Google Cloud Run)}\\[2pt]
     \texttt{POST /api/v1/chat/}\\
     Schema validation, cache \& dispatch};

\node[grpbox, fit=(api_main)(api_chat),
      label={[font=\small\bfseries, fill=white, inner sep=2pt,
              rounded corners=2pt]above:{FastAPI Backend (Google Cloud Run)}}]
      (be_box) {};

\node[svc, below=1.55cm of api_main, xshift=-2.2cm] (mcdm)
    {\textbf{Hybrid MCDM Service}\\[3pt]
     SAW + EDAS + CODAS\\
     CRITIC-Derived Weights\\
     15-min Aggregated Bins};

\node[svc, right=0.4cm of mcdm] (rtsi)
    {\textbf{RT-SI Service}\\[3pt]
     EB Stabilisation\\
     Uplift Factors\\
     VRU \& Vehicle Sub-Indices};

\node[chatsvc, right=0.4cm of rtsi] (chatsvc)
    {\textbf{Chat Service}\\[3pt]
     \texttt{get\_safety\_score}\\
     \texttt{get\_component\_breakdown}\\
     \texttt{get\_historical\_baseline}\\
     \texttt{compare\_intersections}\\
     \texttt{get\_trend\_data}\\
     guarded \texttt{run\_sql\_query}};

\node[llm, right=0.4cm of chatsvc] (llm)
    {\textbf{OpenAI}\\GPT-4o\\[3pt]
     Tool-Augmented\\Function Calling};

\node[dbtab, below=1.5cm of mcdm, xshift=2.2cm] (db_rt)
    {\textbf{Real-Time Tables}\\[2pt]
     \texttt{bsm} | \texttt{psm} | \texttt{vehicle-count}\\
     \texttt{vru-count} | \texttt{speed-distribution} | \texttt{safety-event}};

\node[dbtab, right=0.7cm of db_rt] (db_crash)
    {\textbf{Historical Crash Dataset}\\[2pt]
     VDOT Records 2017--2024\\
     Fatal / Injury / PDO --- Spatially Joined};

\node[grpbox, fit=(db_rt)(db_crash),
      label={[font=\small\bfseries, fill=white, inner sep=2pt,
              rounded corners=2pt]above:{Cloud SQL (PostgreSQL + PostGIS, Google Cloud)}}]
      (db_box) {};

\node[ext, below=1.2cm of db_rt] (ext_vcc)
    {\textbf{VCC / Trino Smart Cities Platform}\\
     BSM \& PSM Connected-Vehicle Telemetry};

\node[ext, below=1.2cm of db_crash] (ext_vdot)
    {\textbf{VDOT Traffic Safety Data}\\
     Police-Reported Crashes (2017--2024)};


\draw[arr] (user.south) -- node[right, font=\scriptsize]{HTTPS}
    (p2.north);

\draw[arr] ([xshift=-0.3cm]p1.south) --
    node[right, font=\scriptsize]{HTTPS / REST (5-min poll)}
    ([xshift=-0.3cm]api_main.north);

\draw[arr] (p5.south) -- node[right, font=\scriptsize]{HTTPS / REST}
    (api_chat.north);

\draw[arr] (api_main.south) -| (mcdm.north);
\draw[arr] (api_main.south) -| (rtsi.north);

\draw[arr] (api_chat.south) -- (chatsvc.north);

\draw[<->, thick] (chatsvc.east) --
    node[above, font=\scriptsize]{Tool calls}
    (llm.west);

\draw[arr] (mcdm.south) -|
    node[pos=0.75, right, font=\scriptsize]{SQL}
    ([xshift=-1.0cm]db_rt.north);

\draw[arr] (rtsi.south) -|
    node[pos=0.75, right, font=\scriptsize]{SQL}
    ([xshift=1.0cm]db_rt.north);

\draw[arr] (chatsvc.south) -|
    node[pos=0.75, right, font=\scriptsize]{SQL (tool calls)}
    (db_crash.north);

\draw[darr] (ext_vcc.north) --
    node[right, font=\scriptsize]{ETL / Collector}
    (db_rt.south);
\draw[darr] (ext_vdot.north) --
    node[right, font=\scriptsize]{Import Script}
    (db_crash.south);

\end{tikzpicture}%
}
\caption{Refined cloud-native system architecture of VTTSI. The left pathway shows the
safety-index pipeline: Streamlit analysis pages call the FastAPI REST endpoints, which
dispatch to the Hybrid MCDM Service and RT-SI Service; both services query the shared
Cloud SQL (PostgreSQL + PostGIS) database. The right pathway shows the SafetyChat
module: the natural-language interface page calls \texttt{POST /api/v1/chat/}, which
routes to a tool-augmented Chat Service that invokes four live-data tools via the
OpenAI GPT-4o function-calling API. Dashed arrows indicate external ETL ingestion
from the VCC/Trino connected-vehicle platform and VDOT historical crash records.}
\label{fig:system-architecture}
\end{figure*}
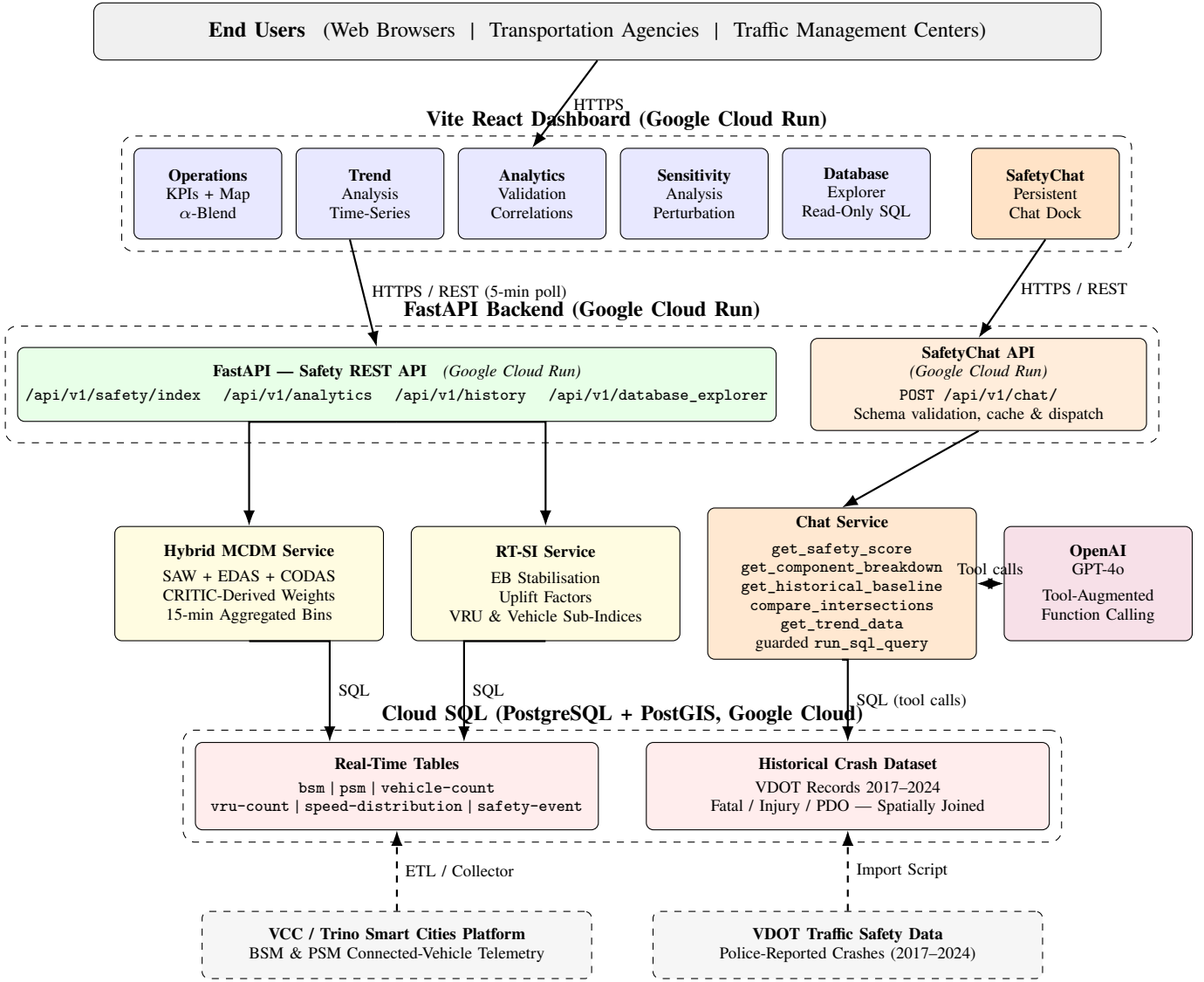

The system adopts a cloud-based client--server architecture with a clear separation between the backend services, frontend visualization layer, and database management components. This modular design ensures scalability, maintainability, and seamless integration with cloud services using CI/CD pipeline.

\subsection{Backend}
The backend is implemented using \textbf{FastAPI}, selected for its asynchronous I/O support and lightweight performance suitable for cloud deployment (see Appendix~\ref{appendix:backend-services} for a summary of backend service components). The backend is currently hosted on \textbf{Google Cloud Run}, allowing automatic scaling and containerized execution. 

The URL for the Backend FastAPI Endpoint is \textbf{\url{https://cs6604-trafficsafety-180117512369.europe-west1.run.app/}}

\noindent The backend is responsible for:
\begin{itemize}
    \item Establishing high-performance connections with the \textbf{PostgreSQL} database using connection pooling.
    \item Fetching raw datasets including Basic Safety Messages (BSM), Personal Safety Messages (PSM), vehicle/VRU counts, and speed distributions.
    \item Executing the \textbf{Hybrid MCDM Service}, which computes safety scores using parallel \textbf{SAW}, \textbf{EDAS}, and \textbf{CODAS} algorithms weighted by the \textbf{CRITIC} method.
    \item Calculating the \textbf{Real-Time Safety Index (RT-SI)} using Empirical Bayes estimates and uplift factors.
    \item Providing RESTful API endpoints that return structured JSON results for visualization.
\end{itemize}

\subsection{Database}
The database layer is deployed on \textbf{Google Cloud SQL} as the central repository for structured crash, traffic, and geospatial data. The database stores both preprocessed 15-minute aggregations and raw datasets to support analytical queries.

\begin{itemize}
    \item \textbf{PostgreSQL with PostGIS}: Chosen for its robust support of geospatial queries, enabling efficient spatial indexing of intersection data and crash locations.
    \item \textbf{Cloud Integration}: The database connects securely to the FastAPI backend via private VPC access to minimize latency and ensure data privacy.
\end{itemize}

\subsection{Frontend System Design}
The frontend is developed using \textbf{Streamlit}, providing a rapid, Python-native framework for building interactive dashboards. It is deployed on \textbf{Google Cloud Run} to enable independent scaling from the backend.

The URL to the Frontend of the Web Application is:   
\textbf{\url{https://safety-index-frontend-180117512369.europe-west1.run.app/}}

\noindent The Streamlit interface offers:
\begin{itemize}
    \item \textbf{Interactive Dashboards}: Displays the final blended Safety Index, severity statistics, and traffic trends using dynamic charts.
    \item \textbf{Geospatial Visualization}: Integrates \texttt{streamlit-folium} to render interactive maps where intersections are color-coded by risk level.
    \item \textbf{Dynamic Score Blending}: Allows users to adjust the $\alpha$ parameter to blend the \textbf{RT-SI} (historical/predictive) and \textbf{MCDM} (real-time anomaly) indices dynamically.
    \item \textbf{Real-Time Data Updates}: Automatically refreshes data every 5 minutes via API polling to reflect the latest traffic conditions.
\end{itemize}

\subsection{System Integration Workflow}
The end-to-end workflow of the cloud system is as follows:
\begin{enumerate}
    \item Users access the \textbf{Streamlit} web app hosted on Google Cloud Run.
    \item Streamlit sends HTTPS requests to the \textbf{FastAPI} backend service.
    \item The backend retrieves relevant records from \textbf{Cloud SQL}, computes the parallel MCDM and RT-SI scores, and returns them as JSON objects.
    \item The frontend receives the raw scores and applies the user-defined weighting ($\alpha$) to compute the final \textbf{Safety Index}.
    \item Results are visualized on the dashboard through interactive maps and time-series charts.
\end{enumerate}

\section{Validation Analysis}
\label{sec:validation}

This section evaluates how the Virginia Tech Transportation Safety Index (VTTSI) behaves
over time and across variables using data from 1--25 November 2025. We focus on three
questions: (1) whether the Real-Time Safety Index (RT--SI), MCDM index, and blended
index exhibit realistic temporal patterns; (2) how strongly they relate to volume,
speed, and surrogate-safety indicators; and (3) how the absence of historical crashes at
Glebe--Potomac affects RT--SI behavior. Other intersections (Birch\_St W\_Broad\_St and
E\_Broad\_St N\_Washinton\_St) show similar patterns and are used as cross-checks.

\begin{figure}
    \centering
    \includegraphics[width=\linewidth]{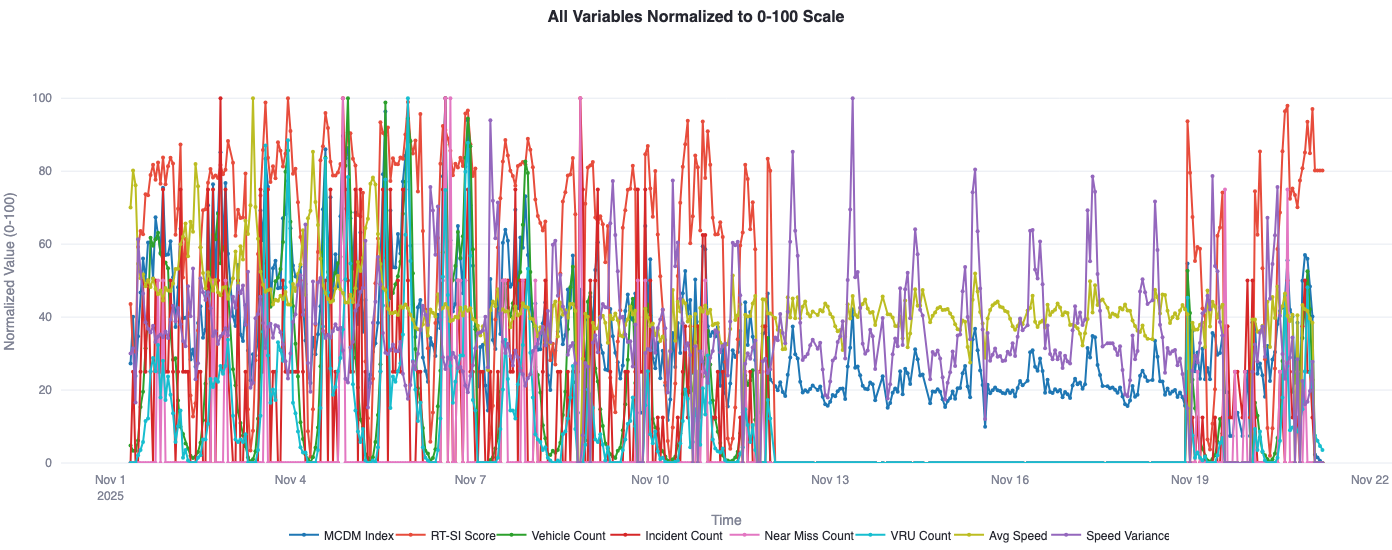}
    \caption{All variables normalized to 0–100 scale for trend analysis at the Glebe--Potomac intersection}
    \label{fig:all-variable-normalized-glebe}
\end{figure}

\begin{figure}
    \centering
    \includegraphics[width=\linewidth]{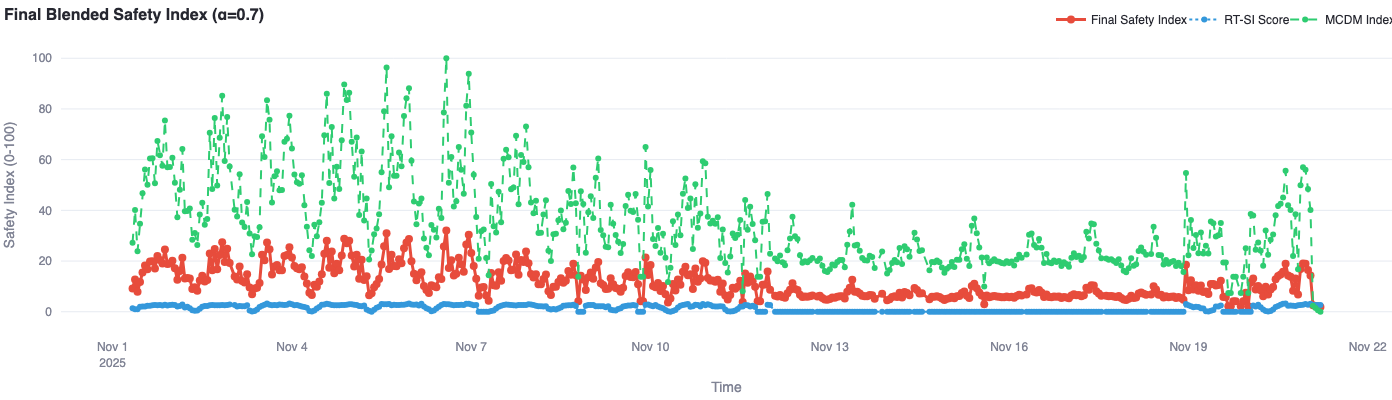}
    \caption{Final blended Safety Index at the Glebe--Potomac intersection}
    \label{fig:final-blended-safety-index-glebe}
\end{figure}

\begin{figure}
    \centering
    \includegraphics[width=\linewidth]{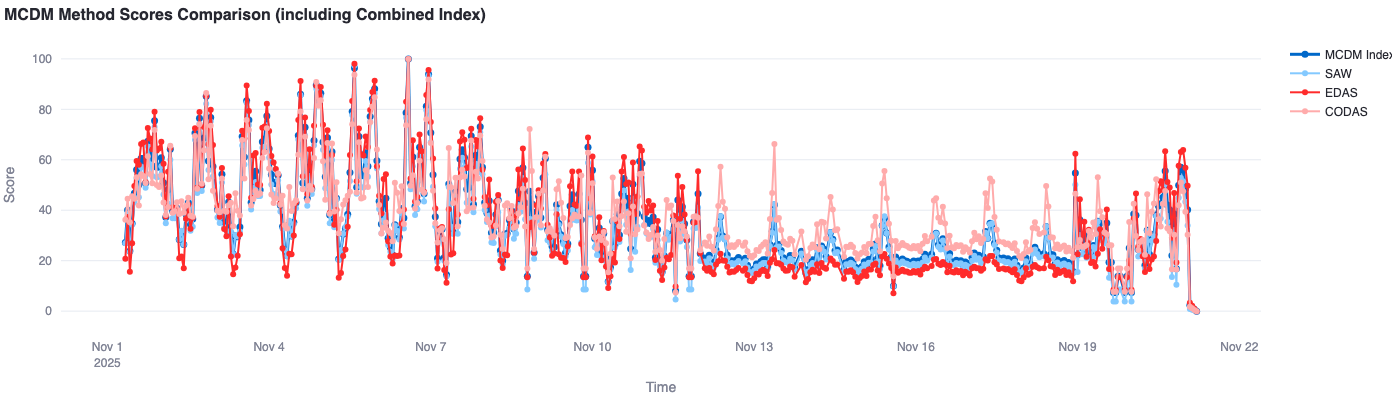}
    \caption{MCDM methods trend comparison at the Glebe--Potomac intersection}
    \label{fig:mcdm-method-scores-comparison-glebe}
\end{figure}

\begin{figure}
    \centering
    \includegraphics[width=\linewidth]{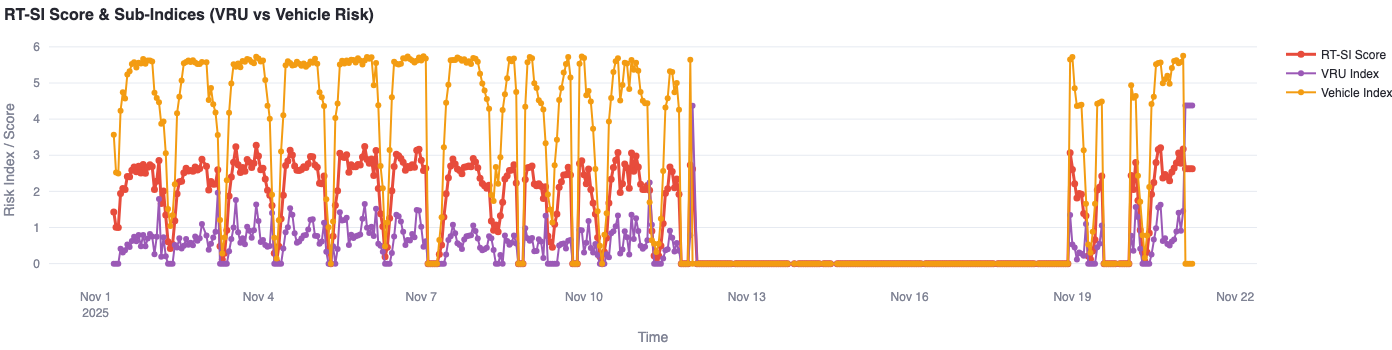}
    \caption{RT--SI scoring methods trend comparison at the Glebe--Potomac intersection}
    \label{fig:rt-si-score-method-comparison-glebe}
\end{figure}

\subsection{Temporal Behavior and Exposure Patterns}

Figure~\ref{fig:all-variable-normalized-glebe} shows a clear diurnal pattern in all
variables: peaks in the morning and afternoon and troughs overnight. Vehicle and VRU
volumes primarily determine the shape of both the MCDM index and the blended Safety
Index, while speed-related terms and incidents perturb that volume-driven baseline.
Similar behavior at Birch and Broad (Appendix figures) confirms that the exposure
signal dominates the daily structure across intersections, consistent with
volume-based safety modeling~\cite{montella2020systemic}.

Figure~\ref{fig:mcdm-method-scores-comparison-glebe} illustrates that the three MCDM
methods (SAW, EDAS, CODAS) produce tightly aligned trends, with occasional small
divergences when one criterion spikes (e.g., speed variance or incidents). This
agreement is expected because all three methods operate on the same normalized
decision matrix and share CRITIC-based weighting.

Figure~\ref{fig:rt-si-score-method-comparison-glebe} shows that RT--SI remains low
(0--6 on the 0--100 scale) at Glebe--Potomac. This is a direct consequence of the
intersection having \emph{no} historical crashes in 2017--2024, so the EB baseline
collapses near the global mean and uplift factors can only slightly raise the score
when turbulence is present~\cite{persaud2007}. In contrast, intersections with
crash history exhibit higher baselines and larger RT--SI ranges, as shown in the
appendix figures.

The blended index (Fig.~\ref{fig:final-blended-safety-index-glebe}) lies between RT--SI
and MCDM, smoothing short-term MCDM spikes while retaining responsiveness. Peaks in the
blended index align with high volumes and elevated speed variance, indicating that the
hybrid design captures both exposure and turbulence.

\subsection{Correlation Structure and Component Consistency}

\begin{figure}
    \centering
    \includegraphics[width=\linewidth]{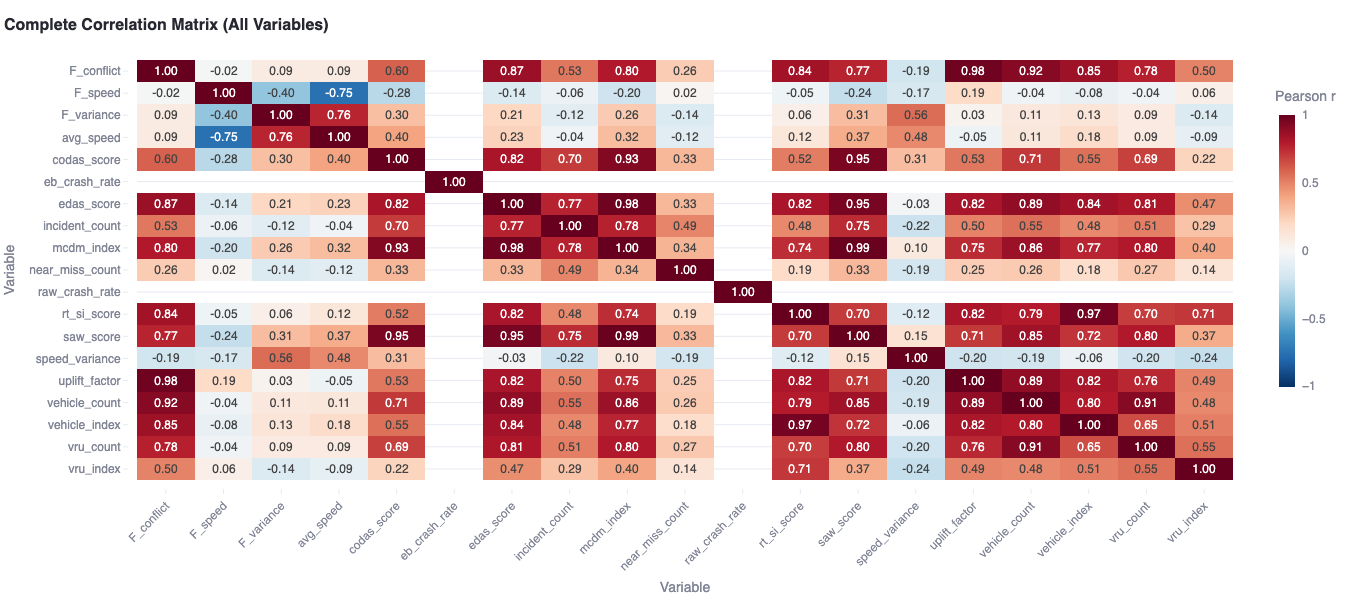}
    \caption{Correlation of variables and safety-score components at the Glebe--Potomac intersection}
    \label{fig:complete-correlation-matrix}
\end{figure}

The Pearson correlation matrix in
Figure~\ref{fig:complete-correlation-matrix} provides a compact view of how components
relate to each other.

\begin{itemize}
    \item \textbf{Internal consistency of MCDM.}  
    SAW, EDAS, and CODAS scores show strong pairwise correlations
    (\(r \approx 0.78\)--\(0.99\)), confirming that all three methods convey a
    consistent operational-risk ordering once CRITIC weights are applied.

    \item \textbf{Exposure dominance.}  
    Vehicle and VRU counts correlate strongly with the MCDM index
    (\(r \approx 0.76\)--\(0.97\)), reflecting that exposure is the primary driver of
    real-time risk in the current formulation, as in prior surrogate-safety and
    conflict-based studies~\cite{schultz2025_surrogate, wang2021highres}.

    \item \textbf{Complementarity of RT--SI and MCDM.}  
    RT--SI and MCDM exhibit a moderate positive correlation, indicating that they
    respond to similar underlying dynamics (volume and turbulence) but with different
    sensitivities. RT--SI is anchored by crash history and uplift bounds; MCDM is
    more sensitive to real-time fluctuations. This justifies treating them as
    complementary views rather than redundant scores~\cite{montella2020systemic}.
\end{itemize}

\subsection{Missing-Data Behavior and Near-Miss Activity}

\begin{figure}
    \centering
    \includegraphics[width=0.8\linewidth]{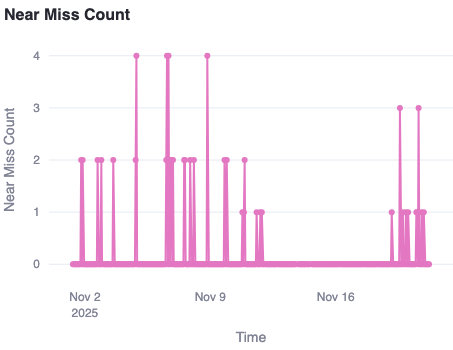}
    \caption{Near-miss count at the Glebe--Potomac intersection}
    \label{fig:near-miss-count}
\end{figure}

During periods with missing or very sparse telemetry, both RT--SI and the MCDM index
collapse to low, stable values: volumes drop to zero, uplift factors vanish, and the
blended index becomes flat. This matches expected behavior for surrogate-safety models
under sensor dropout~\cite{gettman2003ssam} and prevents spurious risk spikes when the
system lacks data.

Near-miss counts (Figure~\ref{fig:near-miss-count}) remain small (typically 0--2 per
15-minute bin) but nonzero, indicating that the sensing pipeline is detecting
conflict-like events at a crash-free location. Because the magnitude is low, the
numerical impact on RT--SI and MCDM is modest, yet the events provide qualitatively
useful evidence that the system is capturing meaningful micro-conflicts, in line with
surrogate-safety practice~\cite{schultz2025_surrogate, gettman2003ssam}.

\subsection{Validation Summary}

Overall, the validation results show that:
\begin{itemize}
    \item the indices follow realistic diurnal patterns driven by exposure,
    \item MCDM methods are internally consistent and strongly tied to volumes and
          turbulence,
    \item RT--SI behaves as an EB-stabilized crash-based index, remaining low at a
          crash-free site but responsive to uplift, and
    \item the blended index provides a stable yet responsive safety signal that
          behaves coherently under missing data.
\end{itemize}
These behaviors match theoretical expectations and prior work on hybrid
crash–surrogate frameworks~\cite{persaud2007, montella2020systemic, schultz2025_surrogate}.

\section{Sensitivity Analysis of RT--SI Parameters}
\label{sec:sensitivity}

To assess robustness of the Real-Time Safety Index (RT--SI) to parameter uncertainty,
we conducted a sensitivity analysis in which all RT--SI parameters were jointly
perturbed by $\pm 25\%$ relative to their baseline values. For each experiment, we
sampled a single parameter vector from a uniform distribution on this range, recomputed
RT--SI for 1--25 November at Glebe--Potomac, and repeated the process for 50 runs.
We then compared each perturbed trajectory to the baseline RT--SI.

\begin{figure}
    \centering
    \includegraphics[width=\linewidth]{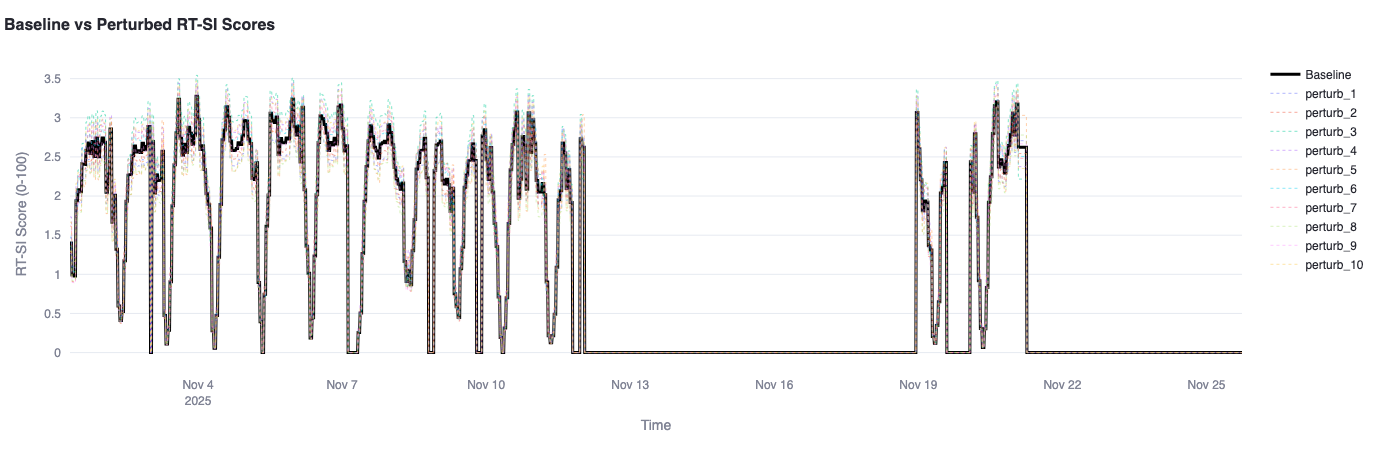}
    \caption{Comparison of perturbed RT--SI trajectories with the baseline at the Glebe--Potomac intersection}
    \label{fig:perturbed-rt-si-comparison-glebe}
\end{figure}

\subsection{Trend Stability and Magnitude of Deviations}

Figure~\ref{fig:perturbed-rt-si-comparison-glebe} shows the baseline RT--SI and a subset
of perturbed trajectories. Even with joint $\pm 25\%$ perturbations, all curves exhibit
the same diurnal structure: morning and afternoon peaks, low nighttime values, and flat
regions when data are missing. Perturbations primarily affect amplitude, not the
location or ordering of peaks and troughs. This indicates that the RT--SI formulation is
structurally stable and dominated by exposure and turbulence signals rather than any
single parameter choice~\cite{schultz2025_surrogate}.

\begin{figure}
    \centering
    \includegraphics[width=\linewidth]{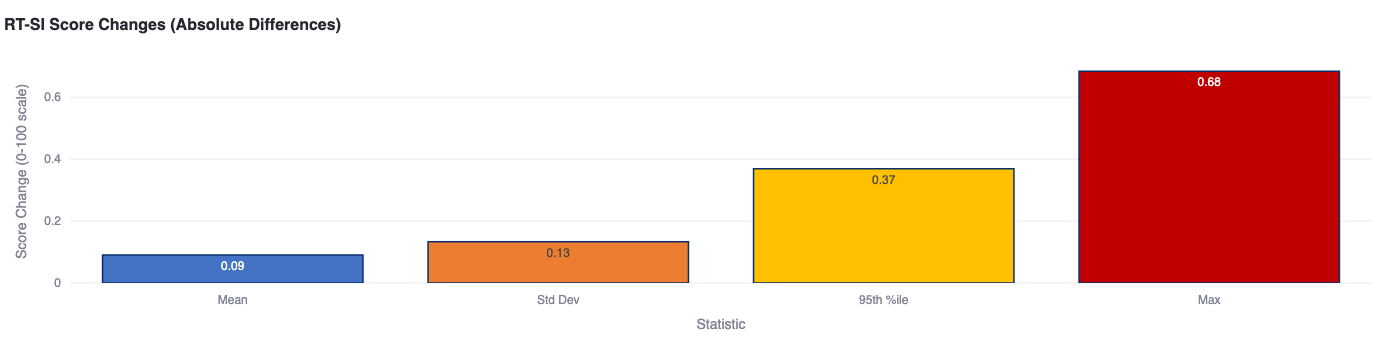}
    \caption{Distribution of absolute deviations between baseline and perturbed RT--SI at the Glebe--Potomac intersection}
    \label{fig:rt-si-changes-distribution-glebe}
\end{figure}

The distribution of absolute deviations (Figure~\ref{fig:rt-si-changes-distribution-glebe})
shows that changes are numerically small:
mean absolute deviation $\approx 0.09$, 95th percentile $\approx 0.37$, and maximum
$\approx 0.68$ on a 0--100 scale. Since RT--SI at this crash-free intersection lies
roughly between 0 and 3.5, these deviations correspond to modest relative changes and do
not alter qualitative conclusions. Such low sensitivity is consistent with EB-based
hybrid models, where uplift terms are bounded~\cite{persaud2007}.

\subsection{Relative Importance of Parameters}

\begin{figure}
    \centering
    \includegraphics[width=\linewidth]{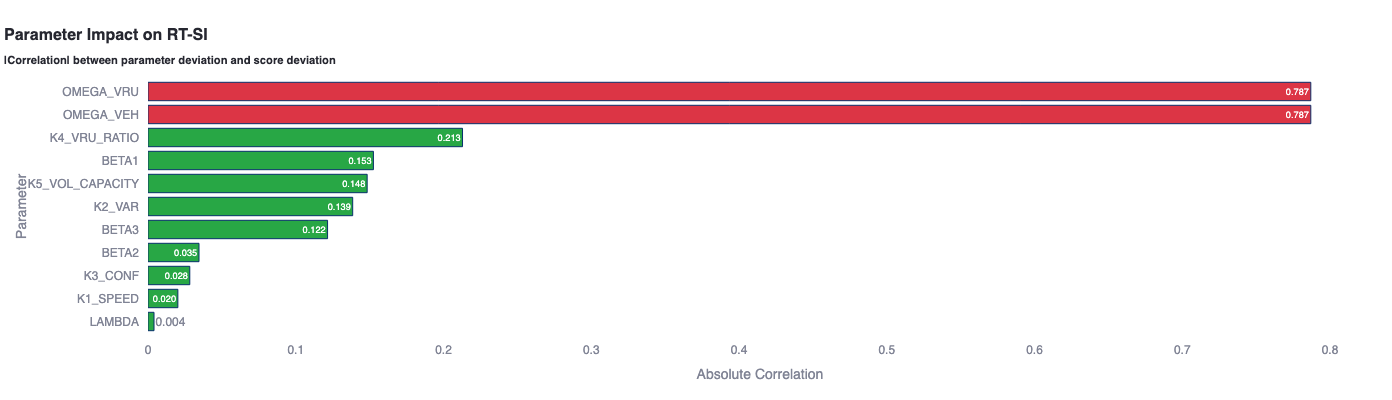}
    \caption{Correlation between parameter perturbations and RT--SI deviations at the Glebe--Potomac intersection}
    \label{fig:weight-importance-rt-si-glebe}
\end{figure}

Figure~\ref{fig:weight-importance-rt-si-glebe} summarizes correlations between parameter
perturbations and RT--SI deviations. Three patterns emerge:

\begin{itemize}
    \item \textbf{VRU and vehicle blending weights dominate.}  
    The VRU and vehicle blending parameters
    $(\omega_{\mathrm{VRU}}, \omega_{\mathrm{VEH}})$ have the strongest correlations
    with score deviations, which is expected because the final RT--SI is a direct
    weighted sum of these sub-indices.

    \item \textbf{Crash-history parameters are weak at crash-free sites.}  
    At Glebe--Potomac, where there are no historical crashes, changes in the EB
    shrinkage parameter and severity weights have minimal effect: the EB baseline is
    essentially the global mean. This mirrors behavior reported for low-crash or
    crash-free locations in the safety literature~\cite{montella2020systemic}.

    \item \textbf{Uplift coefficients are bounded and modest.}  
    Coefficients for speed deficit, variance, and conflict uplift terms have weaker
    influence because speed distributions were relatively stable and conflict volumes
    low, and because each uplift term is explicitly capped in the formulation. This
    bounding prevents extreme scores under noisy conditions~\cite{gettman2003ssam}.
\end{itemize}

\subsection{Robustness Summary}

Taken together, the sensitivity experiments indicate that:

\begin{itemize}
    \item RT--SI preserves its temporal structure under joint parameter perturbations,
    \item score deviations are small even under relatively aggressive $\pm 25\%$ changes,
    \item most sensitivity is tied to explicit policy weights (VRU vs.\ vehicle) rather
          than hidden coefficients, and
    \item the model behaves as intended at a crash-free site, with EB parameters playing
          a limited role and uplift parameters bounded.
\end{itemize}

These findings support using the current RT--SI parameterization for real-time safety
monitoring and suggest that future refinement can focus on policy weights and site
segmentation rather than on fine-tuning the full parameter vector.

\section{Results and Discussion}

This section summarizes the observed behavior of the Virginia Tech Transportation Safety
Index (VTTSI) under real-world intersection data, evaluates the stability and interpretability
of the Safety Index, and discusses insights that emerged during system validation.

\subsection{Temporal Behavior of the Real-Time Safety Index}

Across all test intersections, the crash-based RT--SI exhibited intuitive temporal structure,
with risk increasing during peak periods and decreasing during low-volume periods.
Elevated uplift factors corresponded with identifiable operational anomalies such as 
speed turbulence, VRU presence, or clusters of safety-event detections. These patterns
demonstrate that the hybrid EB+uplift formulation can represent meaningful short-term
fluctuations in operational safety that are not visible in static crash-based assessments
\cite{persaud2007, montella2020systemic, schultz2025_surrogate}.

\subsection{Comparison Between RT--SI and MCDM Operational Risk Scores}

The CRITIC-weighted MCDM index generally aligned with RT--SI for periods of high
operational disturbance, confirming that multiple criteria independently signal elevated
risk. However, the MCDM index was more sensitive to short-term variations in volumes
and speeds, whereas RT--SI remained anchored to historical crash risk. Their complementary
strengths suggest that blended or multi-view safety reporting may offer more complete
situational awareness for operators and planners~\cite{Amraji2025CombinedSafetyIndex, montella2020systemic}.

\subsection{Sensitivity Analysis}

Parameter perturbation experiments (\(\pm 25\%\)) showed that RT--SI is
robust to moderate changes in uplift-factor weights, with typical deviations less than one
point on the 0--100 scale. The index becomes more volatile under very low-volume
conditions. These results indicate that the chosen
features capture stable operational signals but also highlight regimes where additional
context (e.g., conflict topology or multimodal interactions) would improve stability
\cite{gettman2003ssam, wang2021highres}.

\subsection{Interpretation of Real-Time Features}

Correlation analyses reveal that speed variance, safety-event frequency, and VRU
exposure exhibit the strongest association with elevated safety index values. Volume
effects are context dependent: some intersections display increased turbulence under high
flow, while others exhibit risk spikes under moderate flow with heavy VRU activity.
These findings reinforce the need for multimodal modeling and suggest that flat feature
representations may obscure relational effects (e.g., geometry-driven conflict points)
\cite{wang2021highres, schultz2025_surrogate, gettman2003ssam}.

\subsection{Emerging Limitations Observed During Analysis}

The validation process exposed several structural limitations:

\begin{itemize}
    \item Flat feature vectors cannot represent lane-level geometry, conflict relationships,
          or trajectory context.
    \item Normalization rules require site-specific tuning due to heterogeneous operating
          conditions.
    \item VRU under-penetration in PSM data reduces multimodal accuracy.
    \item Safety-event detection is sensitive to missing or inconsistent telemetry.
\end{itemize}

These limitations directly motivated the development of the Agentic Traffic Safety
Knowledge Graph (TS--KG) described in the following section. The observed gaps
make clear that explainability, causal inference, and contextual reasoning require a 
relational representation beyond what tabular features can capture.

\subsection{Answers to Research Questions}

\textbf{RQ1:} The project confirms that historical crash risk and real-time telemetry can be
combined into a unified, interpretable Safety Index.

\textbf{RQ2:} Speed variance, VRU exposure, and safety events exert the most influence on
short-term risk; the index is stable under moderate parameter variation.

\textbf{RQ3:} The RT--SI and MCDM indices show consistent temporal patterns, with useful
complementary sensitivities.

\textbf{RQ4:} The project demonstrates that flat-feature limitations constrain causal inference
and semantic clarity, motivating a move toward an agentic knowledge-graph framework.

\section{Limitations and Future Work}
\label{sec:future-work}

The Virginia Transportation Safety Index (VTTSI) demonstrates the feasibility of
combining historical crash trends with real-time connected-vehicle telemetry, but several
data, modeling, and system limitations influence accuracy, scalability, and
interpretability. These limitations also motivate a clear roadmap for future work.

\subsection{Data and Sensing Limitations}

The current deployment depends heavily on BSM/PSM penetration and the coverage of the
instrumented intersections. PSM adoption remains low, underrepresenting VRUs and
reducing the strength of VRU-related signals. Safety-event detections can be affected by
camera occlusion, blind spots, and classifier errors, and intermittent telemetry produces
flat or low-confidence RT--SI and MCDM values. Crash geolocation from VDOT introduces
spatial ambiguity near intersection boundaries.

\textbf{Future Direction: Multi-Source Plug-In Architecture.}
A unified data-ingestion layer will allow intersections to incorporate additional
modalities—video analytics, LiDAR/CV2X perception, weather feeds, road-surface
conditions, land-use context, and crowdsourced hazard reports. This architecture will
let richer sites contribute more features while still supporting minimal deployments.

\subsection{Limitations of Flat Numeric Representations}

VTTSI currently models intersections using independent, tabular features (volumes, speed
variance, incidents). This loses relational structure such as lane geometry, conflict
topology, multimodal trajectories, and temporal motifs. As a result, causal explanations
(e.g., geometric constraints, recurrent conflict patterns, land-use influences) cannot be
expressed. CRITIC weights also shift based on the last 24 hours, complicating
cross-day comparisons, and at crash-free sites the EB baseline collapses toward zero,
limiting interpretability.

\textbf{Future Direction: Traffic Safety Knowledge Graph (TS--KG).}
A knowledge-graph representation~\cite{cusati2025agentickg} will encode intersections,
approaches, lanes, VRUs, vehicles, trajectories, conflicts, and environmental context as
relational entities. Graph-derived features (motifs, centrality, path risk) will enhance
semantic richness and reduce dependence on flat features.

\subsection{Limits of Numeric-Only Reasoning}

The present model produces numerical scores without semantic explanation or uncertainty
reasoning. It cannot distinguish whether uplift arises from behavior, geometry, weather,
or sensor noise, nor can it trace a chain of evidence for operators.

\textbf{Future Direction: Multi-Agent Reasoning Layer.}
Agents will attach micro-events to the TS--KG, identify recurrent patterns, infer
cause–effect structures, and generate human-readable safety narratives. This will enable
explainable lift factors and risk attribution.

\subsection{Operational and Deployment Constraints}

The current implementation uses a fixed blend parameter $\alpha$, does not offer
context-aware tuning, and has been tested at only a small number of intersections.
Scalability, latency, and reliability under heavy load still require evaluation. VTTSI is
not yet integrated with navigation systems, CAV planners, emergency-response workflows,
or DOT decision-support tools.

\textbf{Future Direction: Ecosystem Integration.}
Future versions will expose open APIs and streaming interfaces to support:
navigation routing, CAV behavior planning, emergency-response prioritization, and agency
network-screening workflows. Automated $\alpha$ selection (rule-based, optimization, or
learning) will tailor the index to different operational needs.

\subsection{Equity and Coverage Gaps}

Sensor coverage is uneven across regions, and intersections in underserved communities
may have lower penetration and poorer telemetry, potentially biasing the index.

\textbf{Future Direction: Equity-Aware Safety Indexing.}
Incorporating demographic, land-use, accessibility, and sensor-coverage indicators will
support fairness-aware calibration and highlight locations where data scarcity masks
risk.

\subsection{Toward a Unified, Agentic Safety Architecture}

Collectively, these limitations point toward an evolution from a purely numeric model to
a relational, agentic, and predictive architecture. Future VTTSI versions will blend
numeric RT--SI and MCDM signals with semantic structures from the TS--KG and reasoning
produced by multi-agent systems, enabling richer explanations, better forecasting, and
more actionable safety intelligence.

\section{Conclusion}
\label{sec:conclusion}

This project developed the Virginia Tech Transportation Safety Index (VTTSI), a
cloud-native, real-time system that integrates severity-weighted Empirical Bayes crash
stabilization with high-frequency connected-vehicle telemetry and a CRITIC-weighted
hybrid MCDM module. Together, these components generate an interpretable 0--100 safety
score every 15 minutes, enabling responsive, data-driven assessment of intersection
conditions.

Through the research questions articulated in this work, we demonstrated that historical
crash history and real-time telemetry can be combined into a unified and
exposure-adjusted Safety Index; that speed variance, VRU exposure, and safety-event
activity form the strongest short-term operational signals; and that the hybrid RT--SI
and MCDM indices provide complementary perspectives on intersection safety. Validation
across multiple intersections confirmed temporal coherence, robustness to parameter
uncertainty, and appropriate sensitivity to operational turbulence and near-miss
activity.

The VTTSI platform provides a practical foundation for continuous safety monitoring and
operational decision support. Its cloud-native architecture and modular service design
make the framework extensible to a statewide deployment. Looking forward, the
Limitations and Future Work section outlines the evolution from flat numerical scoring
toward a relational, agentic, and predictive safety-intelligence architecture.

\section*{Acknowledgment}
The VTTSI system was developed in collaboration with the Virginia Tech
Transportation Institute (VTTI). Generative AI tools, including OpenAI ChatGPT
(GPT-4o) and Anthropic Claude, were used to assist in preparing portions of this
work, including manuscript text, \LaTeX{} formatting, and code snippets. All
AI-generated content was reviewed, verified, and edited by the authors, who take
full responsibility for the accuracy and integrity of the submitted manuscript.

\printbibliography

\vspace{0.5em}

\clearpage
\appendices

\section{Backend Service Descriptions}
\label{appendix:backend-services}

\begin{table}[h]
\centering
\small
\caption{Core Backend Services}
\label{tbl:backend-services}
\begin{tabular}{@{}p{0.28\columnwidth} p{0.63\columnwidth}@{}}
\toprule
\textbf{Service} & \textbf{Description} \\
\midrule
MCDM Safety Index Service
& Computes SAW, EDAS, CODAS, and hybrid MCDM scores. \\

RT--SI Safety Index Service
& Computes Empirical Bayes–stabilized and uplift-adjusted real-time safety index. \\

Telemetry Aggregation Module
& Aggregates BSM, PSM, count tables, and speed-distribution data into aligned 15-minute bins. \\

Safety Event Processor
& Processes safety-event detections and attaches them to analysis windows. \\
\bottomrule
\end{tabular}
\end{table}

\section{Additional Methodology Details}
\label{appendix:methodology}

This appendix expands on the formulations referenced in Section~\ref{sec:methodology},
including parameter tables and the full weight-determination framework used in the
real-time and MCDM safety indices.

\subsection{Parameter Summary}
\begin{table}[ht]
\centering
\begin{tabularx}{\columnwidth}{lX}
\toprule
Parameter & Meaning \\
\midrule
$w_s$ & Severity weights (e.g., Fatal=10, Injury=3, PDO=1) \\
$\lambda$ & EB shrinkage strength \\
$k_1..k_5$ & Scaling constants for speed drop, variance, conflicts, VRU/veh ratios \\
$\beta_1..\beta_3$ & Weights for real-time uplift factors \\
$\omega_{\text{vru}},\omega_{\text{veh}}$ & Weights for VRU vs Vehicle sub-indices \\
$W_E,W_C,W_S$ & Method weights for EDAS, CODAS, SAW in Hybrid index \\
$\alpha$ & Blend factor for Final index (RT vs. MCDM) \\
\bottomrule
\end{tabularx}
\caption{Key tunable parameters of the Safety Index formulas.}
\end{table}

\subsection{Weight Determination Framework}
This section documents the exact weighting scheme implemented in the real-time 
and MCDM components of the Safety Index. All formulas in this section match the 
codebase, including the Empirical Bayes estimator, uplift factors, and CRITIC-based 
objective weighting used in the MCDM module.

\subsubsection{Severity Weights $w_s$}
Severity weighting is applied only in the historical crash-rate computation.  
We adopt fixed policy-driven weights:
\[
w_{\text{fatal}}=10,\qquad 
w_{\text{injury}}=3,\qquad
w_{\text{PDO}}=1,
\]
which match the implementation used to build severity-weighted crash counts.

\subsubsection{Empirical Bayes Shrinkage Parameter $\lambda$}
The historical crash rate for a 15-minute bin is stabilized using a no-exposure 
Empirical Bayes (EB) estimator:
\[
\hat r_i \;=\;
\frac{Y_i + \lambda r_0}{1+\lambda},
\]
where $Y_i$ is the severity-weighted crash count in bin $i$, $r_0$ is the pooled 
mean over 2017--2024, and $\lambda$ is selected using temporal cross-validation.
A grid search over $\lambda\in\{0.1,\ldots,100000\}$ minimizes Poisson negative 
log-likelihood when predicting 2025 outcomes.  
The resulting parameters are:
\[
\lambda^\star = 100000,\qquad r_0 = 3.365.
\]

\subsubsection{Real-Time Uplift Factors $(F_{\text{speed}},F_{\text{var}},F_{\text{conf}})$}
The real-time Safety Index multiplies the EB baseline by three bounded uplift 
factors derived from speed, turbulence, and VRU--vehicle conflict exposure.

\paragraph{Speed deficit uplift.}
Let $\mathrm{FFS}$ denote the free-flow speed and $v$ the observed mean speed:
\[
F_{\text{speed}}
=
\min\!\left(1,\;
1.5\,\frac{\mathrm{FFS}-v}{\mathrm{FFS}}
\right).
\]

\paragraph{Speed variance uplift.}
Let $\sigma^2$ be the variance of speed in the current bin:
\[
F_{\text{var}}
=
\min\!\left(1,\;
1.0\,\frac{\sqrt{\sigma^2}}{v}
\right).
\]

\paragraph{VRU conflict uplift.}
Let $T$ be turning volume and $V_{\mathrm{vru}}$ be VRU count:
\[
F_{\text{conf}}
=
\min\!\left(1,\;
0.5\,\frac{T\,V_{\mathrm{vru}}}{1000}
\right).
\]

\paragraph{Combined real-time multiplier.}
The code uses fixed coefficients $(\beta_1,\beta_2,\beta_3)=(0.3,0.3,0.4)$:
\[
U
=
1
+
0.3F_{\text{speed}}
+
0.3F_{\text{var}}
+
0.4F_{\text{conf}}.
\]

\subsubsection{VRU--Vehicle Index Blend $\omega$}
The real-time index blends VRU-specific and vehicle-specific sub-indices using 
fixed weights:
\[
\mathrm{SI}_{\mathrm{RT}}
=
0.6\,\mathrm{SI}_{\mathrm{VRU}}
+
0.4\,\mathrm{SI}_{\mathrm{VEH}}.
\]
These values are policy parameters, not learned.

\subsubsection{MCDM Criterion Weights $w_j$ (CRITIC Only)}
For the MCDM component, each 15-minute intersection--time pair is treated as an 
alternative, and the following five real-time criteria form the decision matrix:
\begin{equation}
\begin{aligned}
\mathcal{C} = \{\, 
&\text{vehicle\_count},\ \text{vru\_count},\ \text{avg\_speed},\\
&\text{speed\_variance},\ \text{incident\_count}
\,\}.
\end{aligned}
\end{equation}

After min--max normalization, CRITIC weighting is applied:
\[
C_j
=
\sigma_j
\sum_{k}(1-\rho_{jk}),\qquad
w_j
=
\frac{C_j}{\sum_j C_j},
\]
where $\sigma_j$ is the standard deviation of criterion $j$ and $\rho_{jk}$ is 
the correlation between criteria $j$ and $k$ across all intersection--time 
alternatives in the last 24 hours.  
Entropy weighting is not used in the implementation.

\subsubsection{Hybrid Method Weights $W_E,W_C,W_S$ (CRITIC Again)}
After computing SAW, EDAS, and CODAS scores (all normalized to $[0,100]$), these 
three method outputs form a matrix
\[
M = [\mathrm{EDAS}_a,\;\mathrm{CODAS}_a,\;\mathrm{SAW}_a],
\]
from which CRITIC weights are again derived:
\[
W_j
=
\frac{\sigma_j \sum_k(1-\rho_{jk})}
     {\sum_j \sigma_j \sum_k(1-\rho_{jk})},
\qquad 
j\in\{\mathrm{EDAS},\mathrm{CODAS},\mathrm{SAW}\}.
\]
Method weights are recalculated dynamically for each evaluation window.

The hybrid MCDM index is the weighted sum:
\[
\mathrm{SI}^{\mathrm{MCDM}}_{i,t}
=
W_E \,\mathrm{EDAS}_{i,t}
+
W_C \,\mathrm{CODAS}_{i,t}
+
W_S \,\mathrm{SAW}_{i,t}.
\]

\subsubsection{Final Blend}
The released backend computes both the real-time (RT--SI) and MCDM indices and
combines them into the blended Final Safety Index of
Eq.~\eqref{eq:blended-final},
$\mathrm{SI}^{\text{Final}}_{i,t} = \alpha\,\mathrm{SI}^{\text{RT}}_{i,t} +
(1-\alpha)\,\overline{\mathrm{SI}}^{\text{MCDM}}_i$.
The blend coefficient defaults to $\alpha = 0.7$ and is exposed both as a
request parameter on the safety-index API endpoint and as an interactive
slider in the Streamlit frontend, allowing operators to shift continuously
between the historical-risk pole (RT--SI, $\alpha\!\to\!1$) and the
operational-anomaly pole (MCDM, $\alpha\!\to\!0$). The final index ranges from
0 (very safe) to 100 (very dangerous).

\section{Additional Validation Figures}
\label{appendix:validation}

This appendix contains supplemental figures referenced in
Section~\ref{sec:validation}, including additional intersections (Birch and Broad)
to demonstrate consistency of temporal patterns, score behavior, and model coherence.

\subsubsection{Cross-Validation}
Only the Empirical Bayes shrinkage parameter $\lambda$ uses temporal 
cross-validation:
\begin{enumerate}
\item Use 2017--2024 severity-weighted crash counts as training data.
\item Compute pooled mean $r_0$.
\item For each $\lambda$ in a predefined grid:
    \begin{enumerate}
        \item Compute EB-stabilized rates for training years.
        \item Predict 2025 counts.
        \item Compute Poisson log-loss.
    \end{enumerate}
\item Select $\lambda^\star$ minimizing log-loss; hardcode into real-time service.
\end{enumerate}
Weights for uplift factors, VRU--vehicle blending, and MCDM are not tuned via 
cross-validation.

\section*{Validation additional images from other intersection}
\begin{figure*}
    \centering
    \includegraphics[width=\linewidth]{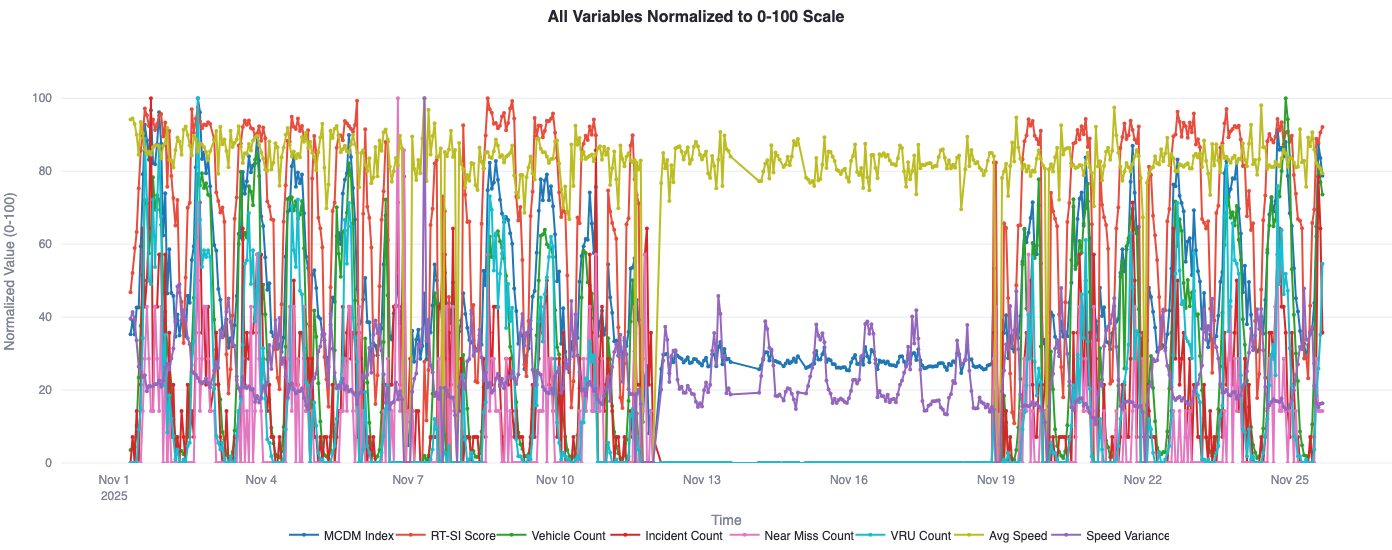}
    \caption{All Variables Normalized to 0-100 scale for trend analysis on  E\_Broad\_St N\_Washinton\_St intersection}
    \label{fig:all-variable-normalized-broad}
\end{figure*}

Results for Birch and Broad intersections exhibit the same behavior and are included in Appendix figures.
Figure~\ref{fig:all-variable-normalized-birch} and figure~\ref{fig:all-variable-normalized-broad} are all variables trend comparisons for each intersection.

\begin{figure*}
    \centering
    \includegraphics[width=\linewidth]{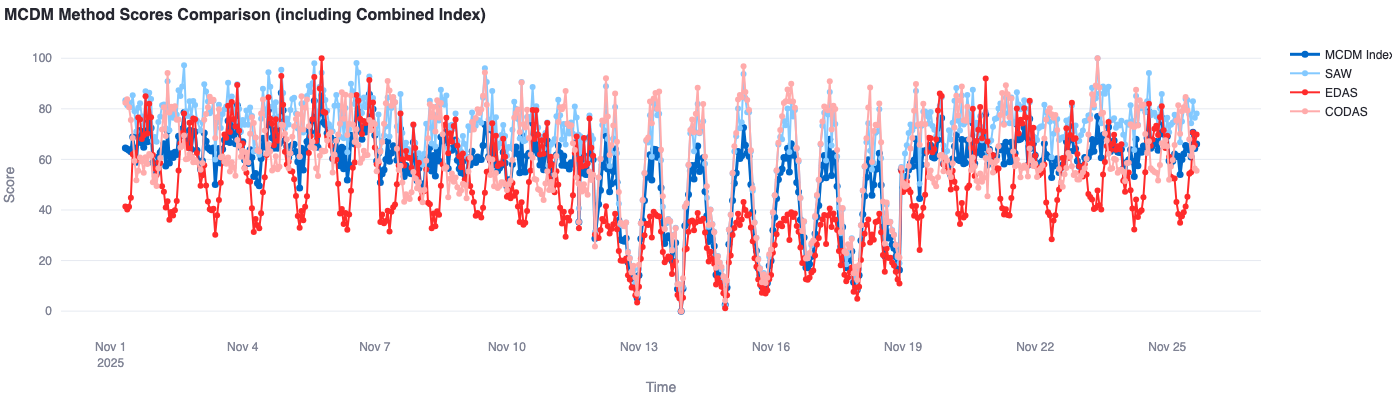}
    \caption{MCDM methods trend comparison on Birch\_St W\_Broad\_St intersection}
    \label{fig:mcdm-method-scores-comparison-birch}
\end{figure*}

\begin{figure*}
    \centering
    \includegraphics[width=\linewidth]{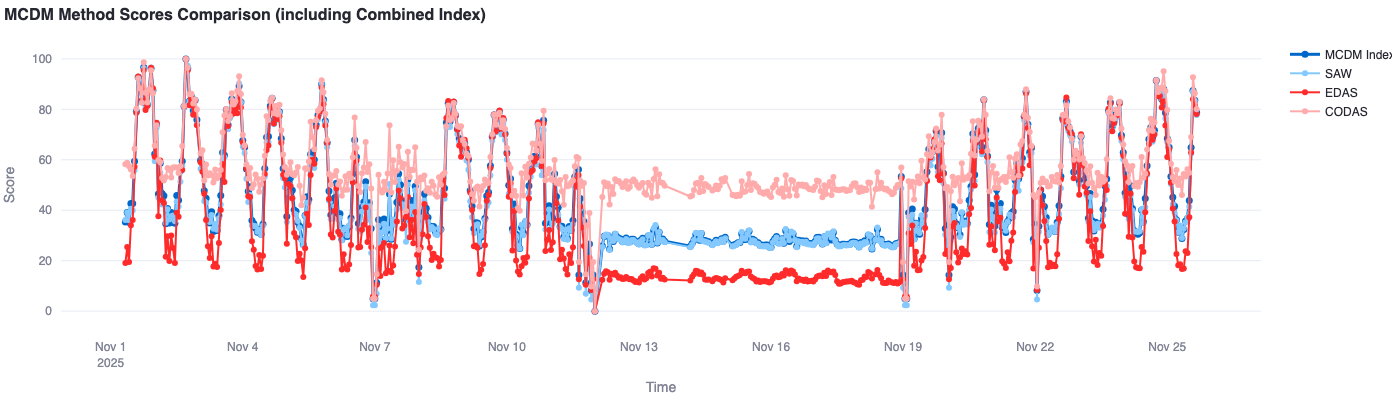}
    \caption{MCDM methods trend comparison on E\_Broad\_St N\_Washinton\_St intersection}
    \label{fig:mcdm-method-scores-comparison-broad}
\end{figure*}

Figure~\ref{fig:mcdm-method-scores-comparison-birch} and figure~\ref{fig:mcdm-method-scores-comparison-broad} are MCDM method and its sub-indices comparison.

\begin{figure*}
    \centering
    \includegraphics[width=\linewidth]{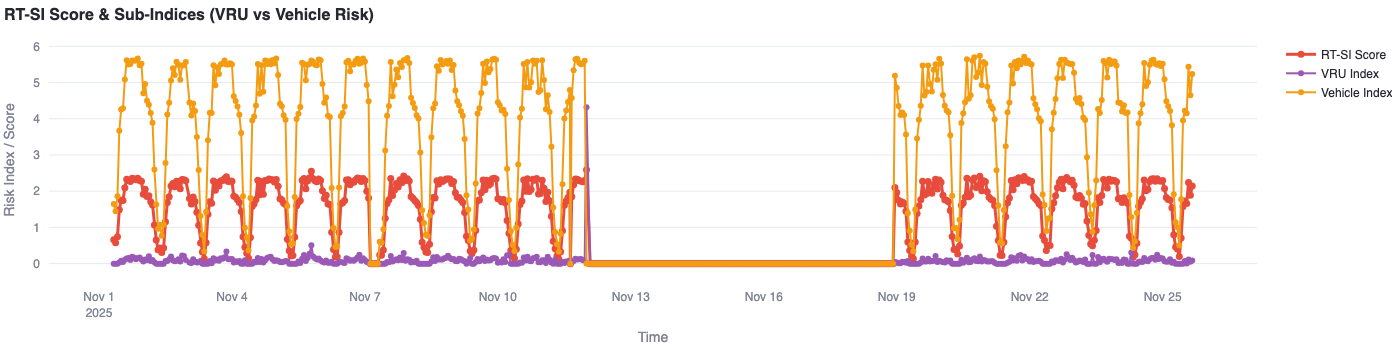}
    \caption{RT-SI scoring methods trend comparison on Birch\_St W\_Broad\_St intersection}
    \label{fig:rt-si-score-method-comparison-birch}
\end{figure*}

\begin{figure*}
    \centering
    \includegraphics[width=\linewidth]{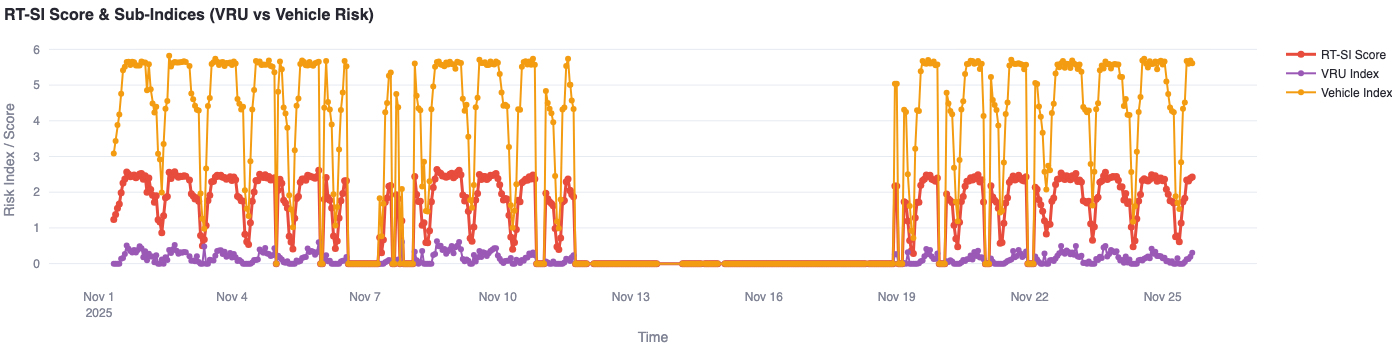}
    \caption{RT-SI scoring methods trend comparison on E\_Broad\_St N\_Washinton\_St intersection}
    \label{fig:rt-si-score-method-comparison-broad}
\end{figure*}

Figure~\ref{fig:rt-si-score-method-comparison-birch} and figure~\ref{fig:rt-si-score-method-comparison-broad} are RT--SI method and its sub-indices comparison.

\begin{figure*}
    \centering
    \includegraphics[width=\linewidth]{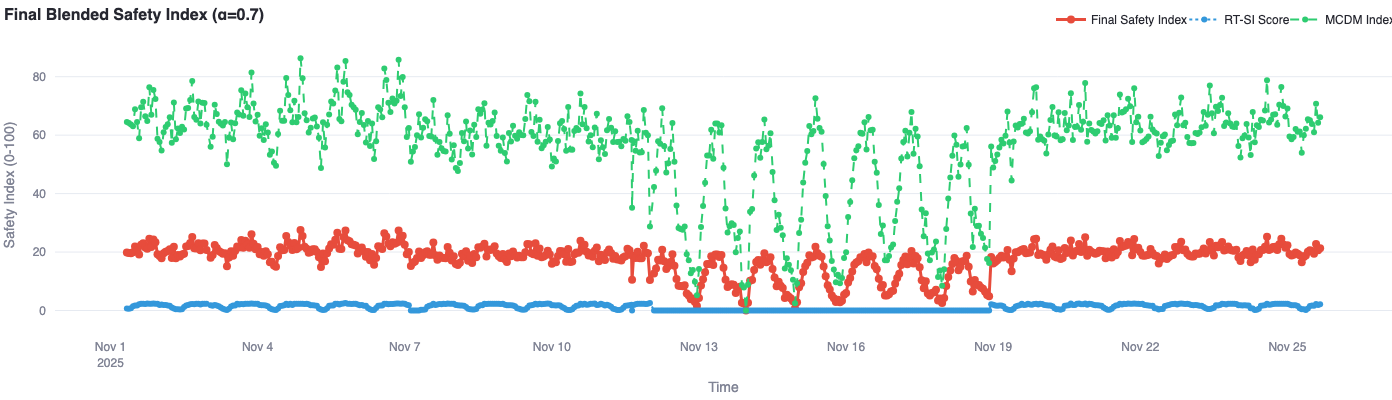}
    \caption{Final Blended Safety Index on Birch\_St W\_Broad\_ Stintersection}
    \label{fig:final-blended-safety-index-birch}
\end{figure*}

\begin{figure*}
    \centering
    \includegraphics[width=\linewidth]{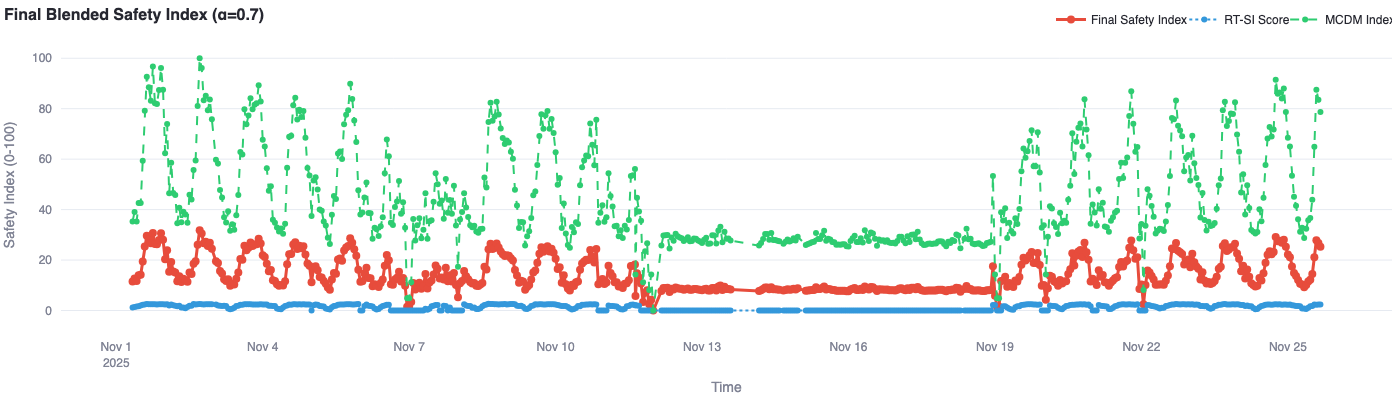}
    \caption{Final Blended Safety Index on E\_Broad\_St N\_Washinton\_St intersection}
    \label{fig:final-blended-safety-index-broad}
\end{figure*}

Figure~\ref{fig:final-blended-safety-index-birch} and figure~\ref{fig:final-blended-safety-index-broad} are how RT--SI and MCDM blended into final index with alpha equal to 0.7.

These images show similar trends and results showed in validation analysis, thus justify further that the conclusion of validation analysis is consistent.

\begin{figure*}
    \centering
    \includegraphics[width=\linewidth]{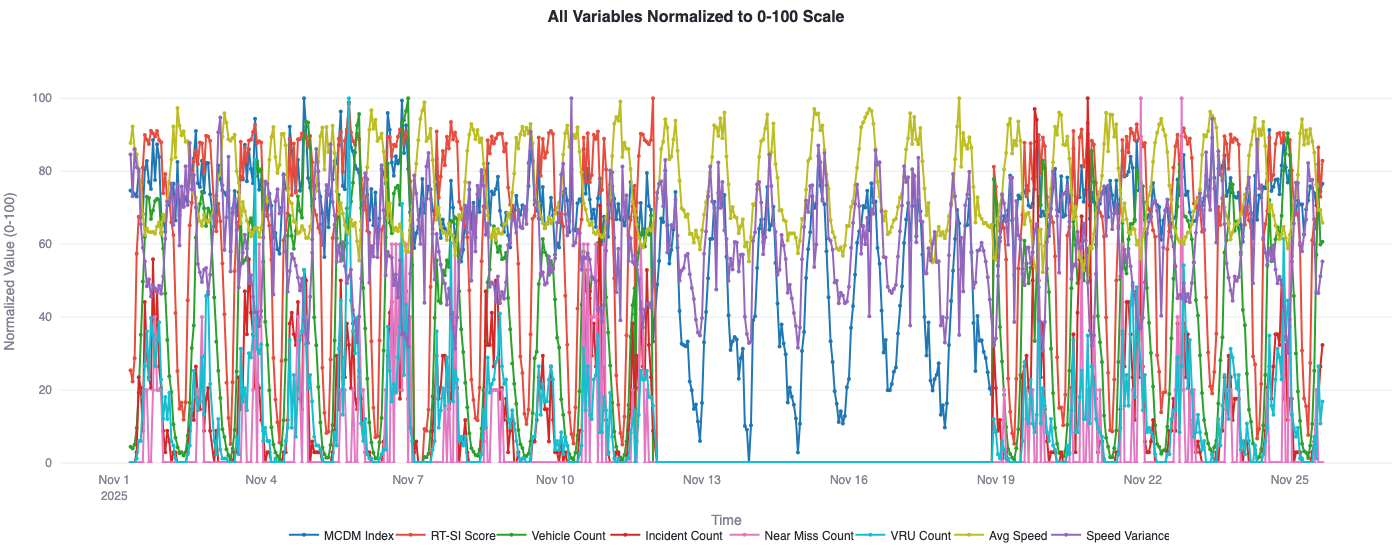}
    \caption{All Variables Normalized to 0-100 scale for trend analysis on Birch\_St W\_Broad\_St intersection}
    \label{fig:all-variable-normalized-birch}
\end{figure*}

\section{Additional Sensitivity Analysis Results}
\label{appendix:sensitivity}

This appendix provides additional plots supporting the sensitivity analysis presented in
Section~\ref{sec:sensitivity}. These results confirm the robustness of the RT--SI under
parameter perturbation across all intersections tested.


Additional information of this sections, including other intersections' results and default setting.

\subsection{Other intersections' results}

\begin{figure*}
    \centering
    \includegraphics[width=\linewidth]{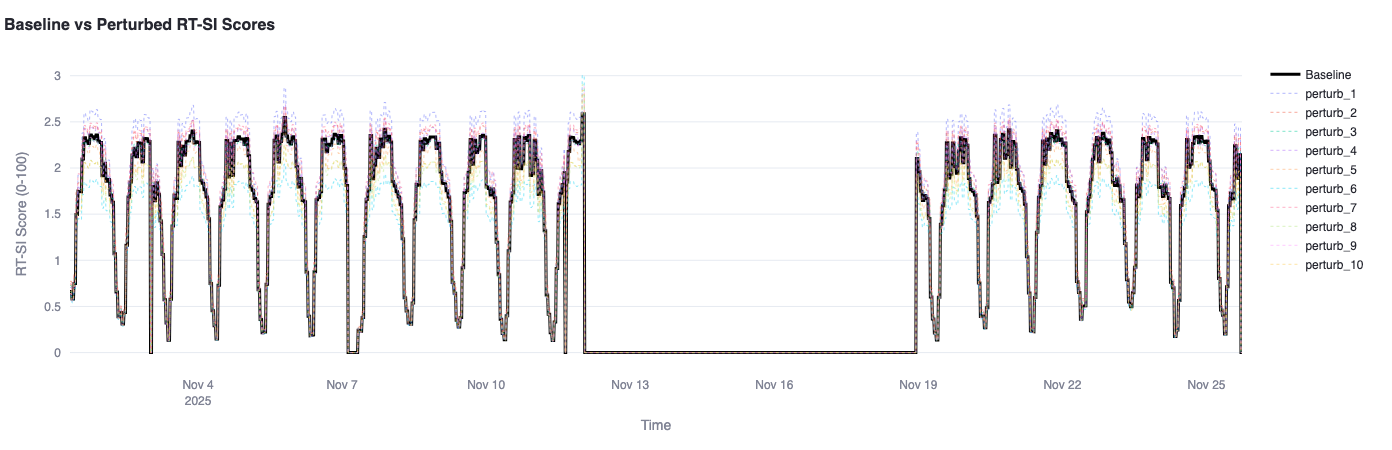}
    \caption{Comparison of perturbations with baseline RT-SI scoring on Birch\_St W\_Broad\_St intersection}
    \label{fig:perturbed-rt-si-comparison-birch}
\end{figure*}

\begin{figure*}
    \centering
    \includegraphics[width=\linewidth]{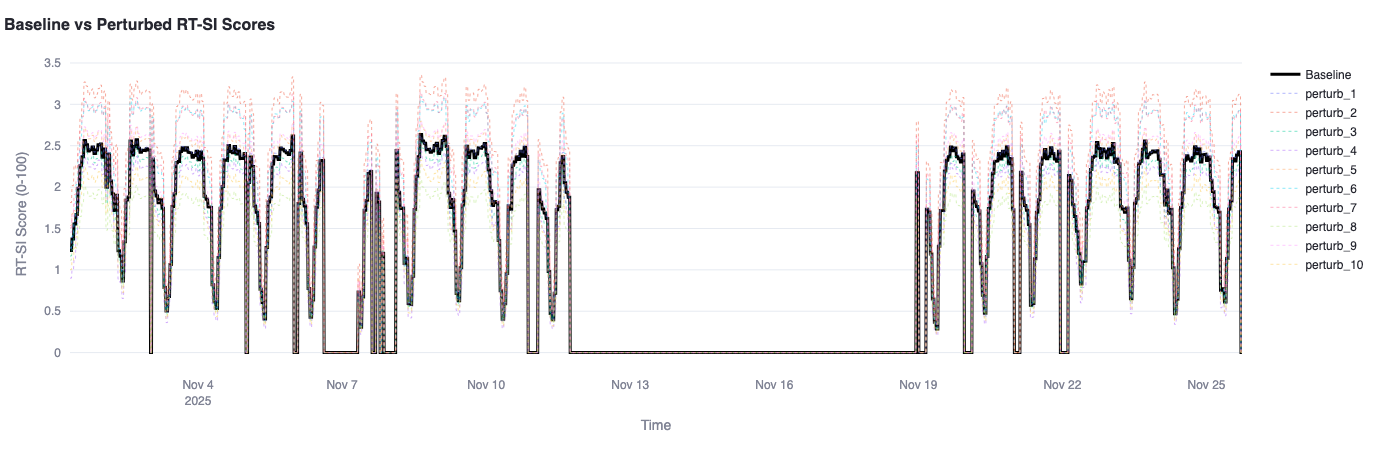}
    \caption{Comparison of perturbations with baseline RT-SI scoring on E\_Broad\_St N\_Washinton\_St intersection}
    \label{fig:perturbed-rt-si-comparison-broad}
\end{figure*}

Figure~\ref{fig:perturbed-rt-si-comparison-birch} and figure~\ref{fig:perturbed-rt-si-comparison-broad} are perturbation results of other intersections.

\begin{figure*}
    \centering
    \includegraphics[width=\linewidth]{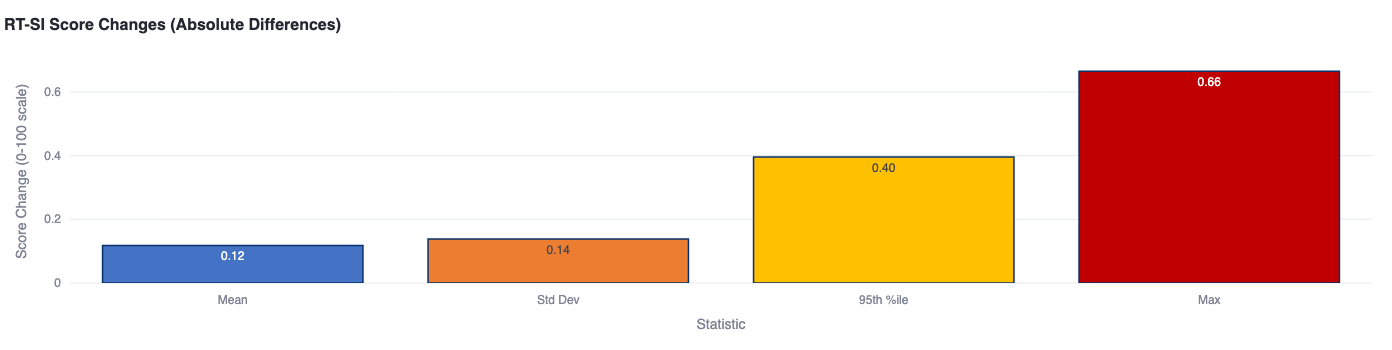}
    \caption{Distribution of absolute deviations between baseline and perturbations on Birch\_St W\_Broad\_St intersection}
    \label{fig:rt-si-changes-distribution-birch}
\end{figure*}

\begin{figure*}
    \centering
    \includegraphics[width=\linewidth]{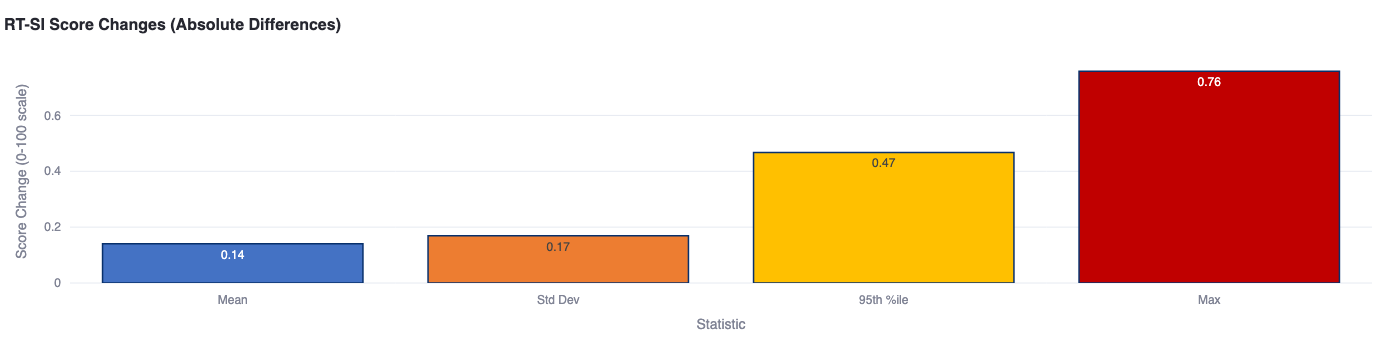}
    \caption{Distribution of absolute deviations between baseline and perturbations on E\_Broad\_St N\_Washinton\_St intersection}
    \label{fig:rt-si-changes-distribution-broad}
\end{figure*}

Figure~\ref{fig:rt-si-changes-distribution-birch} and figure~\ref{fig:rt-si-changes-distribution-broad} are distribution of deviations of perturbations in other intersections.

\begin{figure*}
    \centering
    \includegraphics[width=\linewidth]{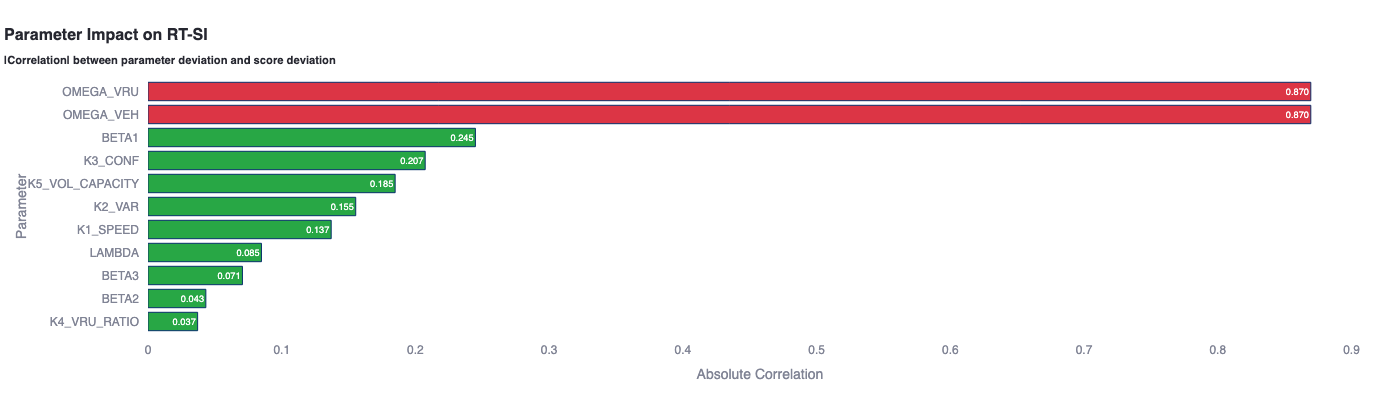}
    \caption{Correlation between parameter
perturbations and RT--SI score deviations on Birch\_St W\_Broad\_St intersection}
    \label{fig:weight-importance-rt-si-birch}
\end{figure*}

\begin{figure*}
    \centering
    \includegraphics[width=\linewidth]{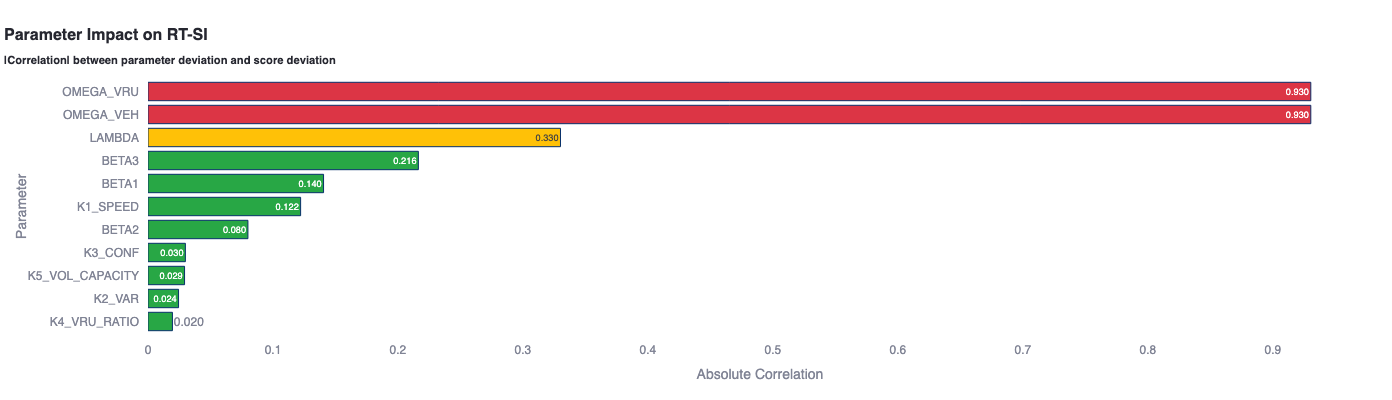}
    \caption{Correlation between parameter
perturbations and RT--SI score deviations on E\_Broad\_St N\_Washinton\_St intersection}
    \label{fig:weight-importance-rt-si-broad}
\end{figure*}

Figure~\ref{fig:weight-importance-rt-si-birch} and figure~\ref{fig:weight-importance-rt-si-broad} shows importances of different weights in other intersections.

These images show similar trends and results showed in sensitivity analysis, thus justify further that the conclusion of sensitivity analysis is consistent.

\subsection{Baseline Parameters}

For reference, the baseline parameters used in the RT--SI model are:

\begin{itemize}
    \item EB parameters:  
    $\lambda = 100{,}000$, \quad $R_0 = 3.365$
    \item Severity weights:  
    $W_{\mathrm{fatal}}=10$, $W_{\mathrm{injury}}=3$, $W_{\mathrm{PDO}}=1$
    \item Uplift scaling:  
    $K1_{\mathrm{speed}}=1.5$, $K2_{\mathrm{var}}=1.0$, 
    $K3_{\mathrm{conf}}=0.5$, $K4_{\mathrm{VRU}}=1.0$, 
    $K5_{\mathrm{vol/cap}}=1.0$
    \item Uplift combination weights:  
    $\beta_1=0.3$, $\beta_2=0.3$, $\beta_3=0.4$
    \item VRU vs vehicle weighting:  
    $\omega_{\mathrm{VRU}}=0.6$, \quad $\omega_{\mathrm{VEH}}=0.4$
    \item Base multiplier: $\gamma=1.0$
    \item Capacity assumption: $500$ vehicles per 15-minute interval
\end{itemize}

\end{document}